\documentclass[journal]{IEEEtran}

\usepackage[T1]{fontenc}
\usepackage{cite}
\usepackage{amsmath,amssymb,amsfonts}
\usepackage{algorithm}
\usepackage{algorithmic}
\usepackage{graphicx}
\usepackage{textcomp}
\usepackage{xcolor}
\usepackage{booktabs}
\usepackage{array}
\usepackage{multirow}
\usepackage{url}
\usepackage{bm}
\usepackage{mathtools}
\usepackage{balance}
\usepackage{tikz}
\usepackage[hidelinks]{hyperref}

\usepackage[top=1.2cm, bottom=1.2cm, left=1.45cm, right=1.45cm]{geometry}

\newcommand{\mcP}{\mathcal{P}}
\newcommand{\mcX}{\mathcal{X}}
\newcommand{\mcU}{\mathcal{U}}
\newcommand{\mcS}{\mathcal{S}}
\newcommand{\mcI}{\mathcal{I}}
\newcommand{\mcK}{\mathcal{K}}
\newcommand{\mcM}{\mathcal{M}}
\newcommand{\mcT}{\mathcal{T}}
\newcommand{\mcN}{\mathcal{N}}
\newcommand{\mcG}{\mathcal{G}}
\newcommand{\mcL}{\mathcal{L}}
\newcommand{\mcR}{\mathcal{R}}
\newcommand{\mcQ}{\mathcal{Q}}
\newcommand{\mcC}{\mathcal{C}}

\newcommand{\mcE}{\mathcal{E}}
\newcommand{\mcY}{\mathcal{Y}}
\newcommand{\mcF}{\mathcal{F}}
\newcommand{\dt}{\Delta t}
\newcommand{\ai}{\mathrm{AI}}

\newcommand{\tr}{\mathrm{tr}}
\newcommand{\infm}{\mathrm{inf}}
\newcommand{\pt}{\mathrm{pt}}
\newcommand{\ft}{\mathrm{ft}}
\newcommand{\lt}{\mathrm{lt}}
\newcommand{\rt}{\mathrm{rt}}

\begin{document}

%\title{Privacy-Preserving Coordinated Operation Scheme Between Power Grid and AI Data Centers}
\title{Privacy-Preserving Coordinated Operation of Power Grids and AI Data Centers: A Checkpoint-Aware Three-Phase Scheme}
\author{
	Ziang~Liu,~\IEEEmembership{Graduate~Student~Member,~IEEE,}
	Ruizhang~Yang,~\IEEEmembership{Graduate~Student~Member,~IEEE,}
	Xin~Cui,~\IEEEmembership{Graduate~Student~Member,~IEEE,}
        Francis~Yunhe~Hou,~\IEEEmembership{Fellow,~IEEE}
%\thanks{
%This work was supported in part by Guangdong Basic and Applied Basic Research Foundation under Grant 2023A1515011942. (\textit{Corresponding author: Francis Yunhe Hou})}
\thanks{Ziang Liu, Ruizhang Yang, Xin Cui and Francis Yunhe Hou are with the Department of Electrical and Electronic Engineering, The University of Hong Kong, Hong Kong (e-mail: liuziang1119@163.com; yhhou@eee.hku.hk).}}

\markboth{IEEE Transactions on Smart Grid,~Vol.~XX, No.~X, Month~2026}%
{Author \MakeLowercase{\textit{et al.}}: Coordination Operation Scheme Between Power Grid and AI Data Center}

\maketitle
\vspace{-2ex}
\begin{abstract}
% The rapid growth of large language model training and serving is pushing AI data centers (AIDCs) toward gigawatt scale, with load characteristics that differ from conventional commercial demand: training and inference workloads can be regulated via dynamic voltage and frequency scaling (DVFS), while periodic model checkpointing produces abrupt power drops and recoveries that can erode operating reserves and stress local transmission capacity. 
The rapid growth of large language model training and serving is driving AI data centers (AIDCs) toward gigawatt scale. Unlike conventional commercial loads, AIDCs possess significant operational flexibility through dynamic voltage and frequency scaling (DVFS) of training and inference workloads, while periodic model checkpointing can induce abrupt power drops and rebounds that erode operating reserves and increase transmission congestion risks.
%Coordinating AIDC operation with grid scheduling under these transients is difficult because grid and AIDC operators are unwilling to disclose proprietary operational data, internal decision-making processes, or operational authority.
Coordinating AIDC operation with grid scheduling under these unique operational characteristics is challenging because grid and AIDC operators are generally unwilling to share proprietary data and decision-making authority.
This paper proposes a hierarchical privacy-preserving coordinated operation scheme between the power grid and AIDCs to address this gap. The proposed scheme contains three phases. 
In Phase I, the grid operator computes a certified inner approximation of the AIDCs security region for subsequent coordination.
%constructs a grid security region for the nodal power trajectories of AIDCs. 
In Phase II, the AIDC operator coordinates training and inference AIDCs to optimize workload allocation within the certified security region and generate power schedules and checkpoint alerts.
%performs internal workload allocation over the security region by coordinating Training AIDCs and Inference AIDCs, and returns the resulting AIDC power schedule and checkpoint timing alerts. 
In Phase III, the grid operator solves a checkpoint-aware two-stage robust optimal power flow (OPF) considering renewable generation and checkpoint uncertainties.
% optimal power flow (OPF)
% , in which renewable generation uncertainty and checkpoint timing uncertainty are simultaneously considered. 
By exchanging only compact interface information, the framework preserves the privacy of both grid and AIDCs, avoids frequent iterative communication, and enables secure coordination with guaranteed feasibility. 
Numerical studies on a modified IEEE 14-bus system and a modified NYISO system demonstrate the effectiveness, robustness, and security of the proposed framework.

\end{abstract}

\vspace{-2ex}
\section{Introduction}
\label{sec:introduction}

\subsection{Background and Motivation}

The rapid development of artificial intelligence (AI), 
% especially large language models,
 has accelerated the expansion of large-scale AI data centers (AIDCs). NERC reports that North American bulk power system peak demand forecasts increase by more than 224 GW in summer and 245 GW in winter over 2026--2035, where data centers account for the majority of this growth~\cite{group2026assessment}. Individual large load facilities are also moving from several hundred megawatts toward gigawatt scale, as illustrated by reported AI training data center projects up to 2 GW~\cite{force2025characteristics}. 
% Compared with conventional commercial loads, AIDCs are spatially concentrated, capacity-intensive, and strongly coupled with computing tasks. Their power consumption depends not only on user demand, but also on the execution of training jobs, inference services, dynamic voltage and frequency scaling (DVFS)~\cite{zhao2023sustainable}, workload routing, and checkpoint operations.
Compared with conventional commercial loads, AIDCs are characterized by significantly larger power demand and more diverse operational characteristics. Different AI workloads exhibit distinct operating behaviors and flexibility mechanisms, such as dynamic voltage and frequency scaling (DVFS), which enables controllable trade-offs between computational throughput and power consumption~\cite{zhao2023sustainable}.

%In large scale pre-training workloads, checkpoint events can introduce significant operational risk for power systems.
%During checkpoint, model parameters and optimizer states are saved for failure recovery, which can create bursty storage traffic and GPU pause time in large clusters. Meta reports Llama 3 training on up to 16K H100 GPUs and checkpoint states ranging from 1 MB to 4 GB per GPU~\cite{meta2024llama3}. From the grid perspective, the transition between active training and checkpoint saving can produce abrupt load drops. 
%NERC reports that such transitions may occur in less than one second, with a measured 50 MW block of a 200 MW AI training data center reaching a fastest ramp rate of 1.9 p.u./s over about 250 ms~\cite{force2025characteristics}. 
%These characteristics make AIDCs both challenging loads and potentially valuable flexible resources for power system operation.

In large-scale AI pre-training, periodic checkpointing is a widely used mechanism for failure recovery, during which massive model parameters and optimizer states are saved to storage. 
For instance, Meta reports that training Llama 3 on 16K H100 GPUs generates checkpoint states ranging from 1 MB to 4 GB per GPU~\cite{meta2024llama3}. 
Writing this enormous volume of data creates bursty storage traffic and forces the entire GPU cluster into a temporary synchronization pause.
% When this internal compute pause occurs, it manifests externally as a severe operational shock to the power grid. 
The rapid transition between active training and checkpoint saving produces abrupt, cliff-like load drops, followed by steep power rebounds when computation resumes. 
NERC reports that such transitions may occur in less than one second, with a measured 50 MW block of a 200 MW AI training data center reaching a fastest ramp rate of 1.9 p.u./s over about 250 ms~\cite{force2025characteristics}. 
These characteristics make AIDCs both challenging loads and potentially valuable flexible resources for power system operation.

From the grid perspective, directly integrating AIDC operation into power system scheduling is challenging. 
% The grid operator needs to ensure power balance, transmission security, reserve adequacy, and robust operation under renewable uncertainty. 
% The AIDC operator controls the DVFS control strategy, cluster-level scheduling, inference routing. A fully centralized model would require extensive data sharing between the power grid and the AIDC operator.
%The AIDC operating can not be directly controlled by grid operators, and the AIDC internal control strategies can greatly influence the external load characteristics. A fully centralized model would require extensive data sharing between the power grid and the AIDC operator.
%, including network information, workload profiles, and proprietary computing models. 
AIDC operation cannot be directly controlled by grid operators, while the internal control strategies of AIDCs can significantly affect their external load characteristics. A fully centralized framework would require extensive sharing of operational data and decision-making authority between the power grid and AIDC operators.
%This assumption is often unrealistic in practice because both entities have privacy, security, and business confidentiality concerns.
Such an assumption is often impractical in real-world applications due to privacy, cybersecurity, and commercial confidentiality concerns on both sides.

% This paper is motivated by the need for a coordination mechanism that is secure for the grid, high efficiency for AIDC, and implementable under limited information exchange. 
%Therefore, a coordination mechanism is needed that simultaneously ensures grid security, preserves AIDC operational efficiency, and can be implemented with limited information exchange.
%The central challenge is to enable the AIDC operator to fully exploit its internal workload flexibility while ensuring that the resulting power trajectories remain acceptable to the grid, and to enable the grid operator to effectively manage the operational risks associated with checkpoint-induced load fluctuations.
%subsequently schedule generation robustly under renewable and checkpoint-related uncertainties.

This creates an operational conflict in modern power system scheduling: while the severe, short-term load drops and rebounds induced by AIDC checkpointing pose operational risk to grid stability, direct grid control over AIDC operations is practically impossible due to data privacy and commercial confidentiality boundaries. Therefore, bridging this gap by establishing a secure, privacy-preserving coordination framework that respects AIDC operational autonomy while ensuring grid reliability has emerged as a pressing necessity.

\vspace{-2ex}
\subsection{Literature Review}

The operation of conventional data centers has been extensively studied. Some studies focus on workload allocation and demand management within data centers, where spatial and temporal workload flexibility is leveraged to reduce electricity costs, improve energy efficiency, and mitigate carbon emissions.
A carbon-aware compute management framework for hyperscale data centers is proposed in~\cite{radovanovic2022carbon}, where temporally flexible workloads are shifted across time through virtual capacity curves to reduce carbon emissions and peak power consumption based on day-ahead carbon intensity forecasts and workload predictions.
Subsequent studies further extended the research by proposing a distributionally robust scheduling framework for geographically distributed data centers, where temporal and spatial workload flexibility were jointly leveraged to achieve carbon- and cost-efficient workload allocation with probabilistic performance guarantees under uncertain compute loads and demand response events~\cite{hall2025carbon}.
More recently, reinforcement-learning-based scheduling strategies have also been investigated for low-carbon data center operation. A multi-objective low-carbon scheduling method based on ensemble deep reinforcement learning was proposed in~\cite{wang2025multi}, where carbon emissions and quality of service were jointly optimized through Pareto-front-based job scheduling using real-time carbon intensity signals and workload information.
A two-stage game-theoretic demand response framework for geo-distributed data centers was proposed in~\cite{wu2025game}, where customer demand was decomposed into regional sub-targets through a potential game and workload allocation was further optimized through an evolutionary game under bounded rationality and incomplete information.
These studies provide useful insight into the controllability and flexibility of computing workloads, but the grid is often represented only through exogenous prices, carbon signals, or simplified capacity limits. Moreover, the operational risks introduced by checkpoint-induced load fluctuations~\cite{force2025characteristics} are generally not considered.

Several studies have explored the coordination between data centers and power systems. A compatible Internet data center load model for demand response was developed in~\cite{chen2020internet}, where geographically distributed data centers were coordinated through workload balancing, batch workload scheduling, and thermal storage to capture their spatial and temporal flexibility. Data centers were also investigated as fast frequency response resources in low-inertia power systems by leveraging delay-tolerant workload control and on-site UPS systems~\cite{al2021potential}. Their frequency support capability was further quantified through a frequency-secured unit commitment framework that co-optimized data center response with generator scheduling and incorporated nonlinear nadir constraints using data-driven linearization~\cite{ren2026grid}. A coordinated bidding strategy jointly optimized workload dispatch, cooling control, electricity purchases, and frequency regulation capacity under workload and regulation-signal uncertainty~\cite{zou2025coordinating}. However, these studies mainly adopt centralized or jointly optimized formulations that model power systems and data centers as an integrated entity. They therefore require extensive information sharing and do not explicitly address data privacy and confidentiality.

%Data centers have also been coordinated with power systems for demand response~\cite{chen2020internet}, fast frequency response~\cite{al2021potential}, frequency-secured generation scheduling~\cite{ren2026grid}, and coordinated energy and regulation bidding~\cite{zou2025coordinating}. These studies mainly adopt centralized or jointly optimized formulations that require extensive information sharing between grid and data center operators. As a result, they generally rely on extensive information sharing and do not explicitly address data privacy and information confidentiality issues.

%To address this limitation, a privacy-preserving coordinated framework was proposed in~\cite{chen2025spatial}, where the coupled DSO–DC problem was decomposed and solved through iterative exchange of limited information between each DSO and its local data center using ADMM.
%
To address this limitation, privacy-preserving coordination methods based on distributed optimization have been investigated. In~\cite{chen2025spatial}, the coupled DSO--DC problem was decomposed and solved using ADMM through iterative exchange of limited information.
A day-ahead co-dispatch framework for hierarchical edge--fog--cloud data centers and distribution networks was developed in~\cite{liu2025synergising}, with the optimization problem solved in a distributed and privacy-preserving manner.
However, practical implementation of such iterative coordination schemes may remain challenging due to communication bandwidth limitations, data security concerns, and the uncertain convergence time and iteration requirements associated with multi-round information exchange.

%To reduce the dependence on sensitive operational data and computationally intensive real-time optimization, a contextual-regression-based coordination mechanism was proposed in~\cite{dvorkin2024agent}, where cost-effective spatial task shifting among geographically distributed data centers was approximated using public contextual signals such as real-time electricity prices.
To reduce the dependence on sensitive operational data, a contextual-regression-based coordination mechanism was proposed in~\cite{dvorkin2024agent}, where spatial task shifting among geographically distributed data centers was approximated using public contextual signals such as real-time electricity prices. The grid model was significantly simplified, with regression-based policies employed to approximate OPF decisions.
% From the perspective of coordinated market participation, geographically distributed data centers were further modeled in~\cite{chen2025spatial} as geographically coupled flexibility providers that enable the coordinated operation of multiple local flexibility markets through cross-regional workload dispatch under uncertainty.

% Most existing formulations model the coupled problem as a centralized or jointly optimized decision-making problem. Such a formulation is mathematically convenient, but it implicitly assumes that the grid operator and the data-center operator are willing to share detailed network, generator, workload, and business information. Distributed or iterative solution methods can reduce the computational burden of the centralized model, but they still require repeated information exchange. In engineering implementation, the number of iterations, convergence time, and communication reliability are difficult to guarantee, especially for day-ahead operational processes with strict time requirements.

The recent transition from conventional data centers to AIDCs further changes the coordination problem. Compared with traditional data centers, AIDCs are driven by heterogeneous training and inference workloads~\cite{he2025freesh}, GPU-cluster synchronization~\cite{kakolyris2025throttll}, DVFS behavior~\cite{kakolyris2024slo}, and checkpoint-induced power variations~\cite{chen2025electricity}. Research on AIDC-grid coordination is still at an early stage. 
Some studies investigate price- or carbon-signal-based mechanisms for guiding AIDC operation~\cite{he2025freesh}, where query routing, GPU parallelism, DVFS, and scheduling policies were jointly optimized across heterogeneous GPU clusters under spatiotemporal variations in carbon intensity and workload patterns.
However such price- or carbon-incentives may be insufficient when electricity cost is small compared with the business value of training progress, model delivery, and service quality. Therefore, a new coordination scheme is needed to reflect both grid-side security requirements and AIDC-side operational autonomy.

The above literature leaves several gaps that motivate this paper:
\begin{itemize}
    \item Existing studies mainly focus on conventional data centers rather than rapidly expanding AIDCs. Their workload composition, load characteristics, and flexibility mechanisms are fundamentally different from those of AIDCs. In particular, checkpointing events in large-scale AI training can induce substantial load fluctuations that are largely overlooked in existing studies and may pose significant risks to power system operation.
    \item Data center studies often optimize workload allocation without explicit grid security concerns, whereas grid operation studies often model data centers as fully controllable loads. Existing literature therefore either neglect the privacy boundary between grid and AIDC operators or require extensive information sharing that may be impractical in real-world deployments.
    %Neither treatment adequately captures the privacy boundary between the power grid and the AIDCs.
   % \item Some studies recognize power fluctuations during AI training, but the analysis is often conducted at the facility or infinite-bus level. The impacts on system-wide economic dispatch, unit commitment, transmission security, and robust scheduling remain insufficiently addressed.
    \item Many existing coordination mechanisms rely on electricity prices or carbon signals to influence data center operation. These incentives may not provide sufficient motivation for AIDC operators to adjust workload schedule. 
    Several studies rely on iterative coordination schemes with repeated information exchange, which may face practical challenges due to communication bandwidth and uncertain convergence times.
\end{itemize}
% These limitations highlight the need for a privacy-preserving coordination framework that simultaneously guarantees grid security, preserves AIDC operational autonomy, and requires only limited information exchange.
These limitations highlight the need for a privacy-preserving coordination framework that guarantees grid security and preserves AIDC operational autonomy without iterative cross-entity communication.
\vspace{-2ex}
\subsection{Contributions and Paper Organization}

This paper develops a hierarchical coordinated operation scheme between the power grid and AIDCs. The main contributions are summarized as follows.

\begin{itemize}
    \item A privacy-preserving three-phase coordination framework is proposed. The framework avoids full data sharing between the power grid and the AIDCs, and only exchanges AIDC nodal power security regions, planned AIDC power trajectories, and checkpoint alert information.
    \item A grid-side security region construction method is developed for multiple AIDC nodes over multiple time periods. The method treats all AIDC nodal powers over the scheduling horizon as a high-dimensional vector and constructs a certified inner V-representation approximation, which is subsequently transmitted to the AIDC operator for workload scheduling.
    \item A workload coordination model for AIDCs is developed, in which large-scale training AIDCs and small-scale inference AIDCs are modeled separately. The model optimizes DVFS decisions and inference workload routing while ensuring that the resulting AIDC power trajectories remain within the grid-certified security region.
    \item A grid-side checkpoint-aware two-stage robust OPF is formulated to validate grid operation under both renewable generation uncertainty and checkpoint timing uncertainty.
\end{itemize}

The remainder of this paper is organized as follows. Section~\ref{sec:overview} introduces the overall coordination framework. Sections~\ref{sec:step1}--\ref{sec:step3} present the three phases of the proposed scheme. Section~\ref{sec:case} describes the case-study design. Section~\ref{sec:conclusion} concludes the paper.

\section{Coordinated Operation Framework}
\label{sec:overview}

\begin{figure}[!b]
    \centering
    \includegraphics[width=\linewidth]{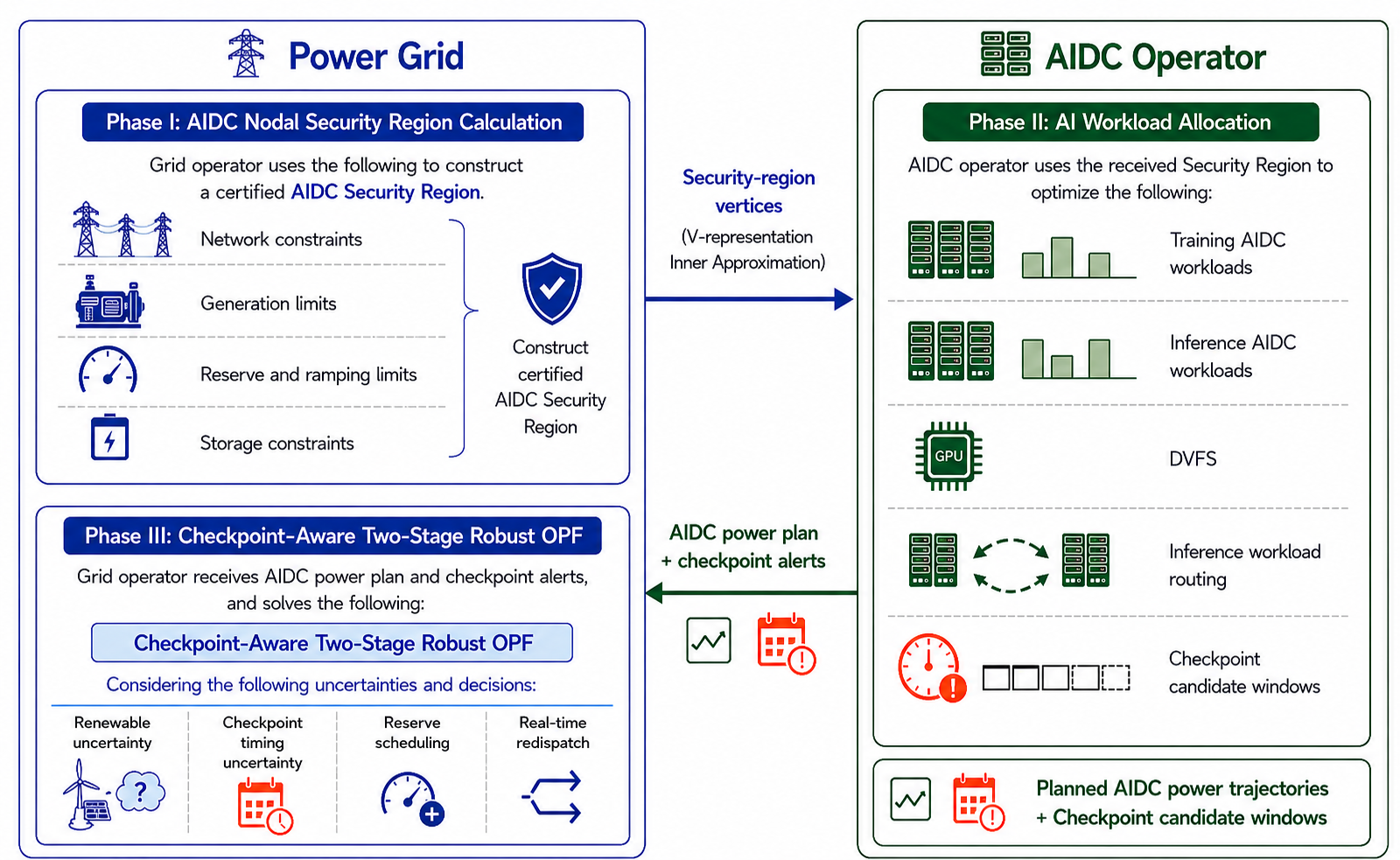}
    \caption{Proposed three-phase coordinated operation framework between the power grid and AIDC operator.}
    \label{fig:framework}
\end{figure}

The proposed coordinated operation framework is demonstrated in Fig.~\ref{fig:framework}. It coordinates two entities with distinct responsibilities and private information: the power grid maintains secure system operation, while the AIDC operator allocates heterogeneous AI workloads, including training jobs, real-time user inference (RT) workload, and latency-tolerant inference (LT) workload.

% Centralized co-optimization would require extensive disclosure of grid-side operating limits and AIDC-side workload data, which is often unacceptable because of privacy, cyber-security, and business-confidentiality concerns. Iterative decomposition can reduce modeling centralization, but frequent cross-entity communication may still hinder practical employment. 
The proposed scheme therefore decomposes the coordination task into three sequential phases, in which each entity independently makes decisions based solely on exchanged interface information, without requiring access to global system information.
\vspace{-2ex}
\subsection{Three-Phase Coordination Framework}

\emph{Phase I, AIDC nodal security region calculation:} A set of certified nodal AIDC power trajectories is constructed by the grid operator.
Each trajectory lies within the high-dimensional security region defined by system-level constraints, including power balance, branch-flow constraints, generator limits, ramping limits, reserve requirements, and storage energy constraints. These trajectories collectively form a V-representation inner approximation of the projected AIDC security region.

\emph{Phase II, AI workload allocation:} The allocation of heterogeneous workloads with spatiotemporal flexibility is optimized by the AIDC operator, while the resulting AIDC power trajectories are kept within the inner-approximated security region. Hence, the AIDC scheduling decisions comply with grid security requirements without requiring access to the full grid model.

\emph{Phase III, checkpoint-aware two-stage robust OPF:} After the AIDC power plan and checkpoint alert information are received by the grid operator, a two-stage robust OPF is solved to validate grid operation under renewable and checkpoint timing uncertainties. The first stage determines generation set points, renewable set points, and reserve schedules. The second stage models real-time redispatch, storage operation, renewable curtailment, load shedding, and network feasibility.
\vspace{-2ex}
\subsection{Data Exchange Between Entities}

Under the proposed scheme, the power grid and the AIDC operator make operational decisions independently. The grid-side problems are solved by the grid operator, while the AI workload allocation problem is solved by the AIDC operator. Coordination is achieved through a compact one-pass interface rather than repeated cross-entity iterations, which substantially reduces communication requirements while preserving the privacy of both parties.

The required cross-entity information exchange is limited to the following two information exchanges:
\begin{itemize}
    \item a set of certified AIDC power trajectory vertices is sent from the grid operator to the AIDC operator;
    \item the optimized AIDC power trajectory and checkpoint candidate windows are sent from the AIDC operator to the grid operator;
\end{itemize}
No detailed internal workload model is required by the grid operator, and no full transmission-network model is required by the AIDC operator. This limited information exchange is one of the main reasons why the proposed method is more practical than a monolithic centralized formulation or a communication-intensive iterative coordination scheme.

% \subsection{Model Representation}

% Let $\mcT=\{1,\ldots,T\}$ denote the scheduling periods with interval length $\dt$. 
% The set of AIDCs is denoted by $\mcI=\mcI^{\tr}\cup\mcI^{\infm}$, where $\mcI^{\tr}$ and $\mcI^{\infm}$ are the training and inference AIDC sets, respectively. The active power consumption of AIDC $i$ at period $t$ is denoted by $p_{i,t}^{\ai}$. 
%The full-horizon AIDC power vector is
%\begin{equation}
%    \bm{x}^{\ai}
%    =
%    \mathrm{col}\{p_{i,t}^{\ai}: i\in\mcI,\ t\in\mcT\}
%    \in \mathbb{R}_+^{|\mcI|T}.
%    \label{eq:ai-vector}
%\end{equation}
% where $\bm{x}^{\ai}$ is the main coordination variable exchanged between the grid-side security-region model and the AIDC-side workload model.

\section{Phase I: AIDC Nodal Security Region Calculation}
\label{sec:step1}

The first task in the proposed framework is to construct a grid operating envelope for AIDC power consumption. From the grid operator's perspective, this envelope should specify admissible multi-period power trajectories at multiple AIDC buses, while avoiding disclosure of detailed grid information. When multiple AIDC buses and multiple time periods are considered jointly, the security region cannot be represented by independent upper and lower bounds. It becomes a high-dimensional polytope in the space of AIDC nodal power trajectories.
Accordingly, a compact and privacy-preserving representation is developed to convert detailed grid operational constraints into information that can be communicated to the AIDC operator. 
% All time periods are modeled jointly, and each transmitted vertex represents a complete AIDC power trajectory over the scheduling horizon. The resulting interface is conservative: if the AIDC operator selects a trajectory inside the communicated security region, that trajectory is guaranteed to be feasible for the grid-side security model.

\subsection{Projected Security Region}

\begin{figure}[!t]
    \centering
    \includegraphics[width=\linewidth]{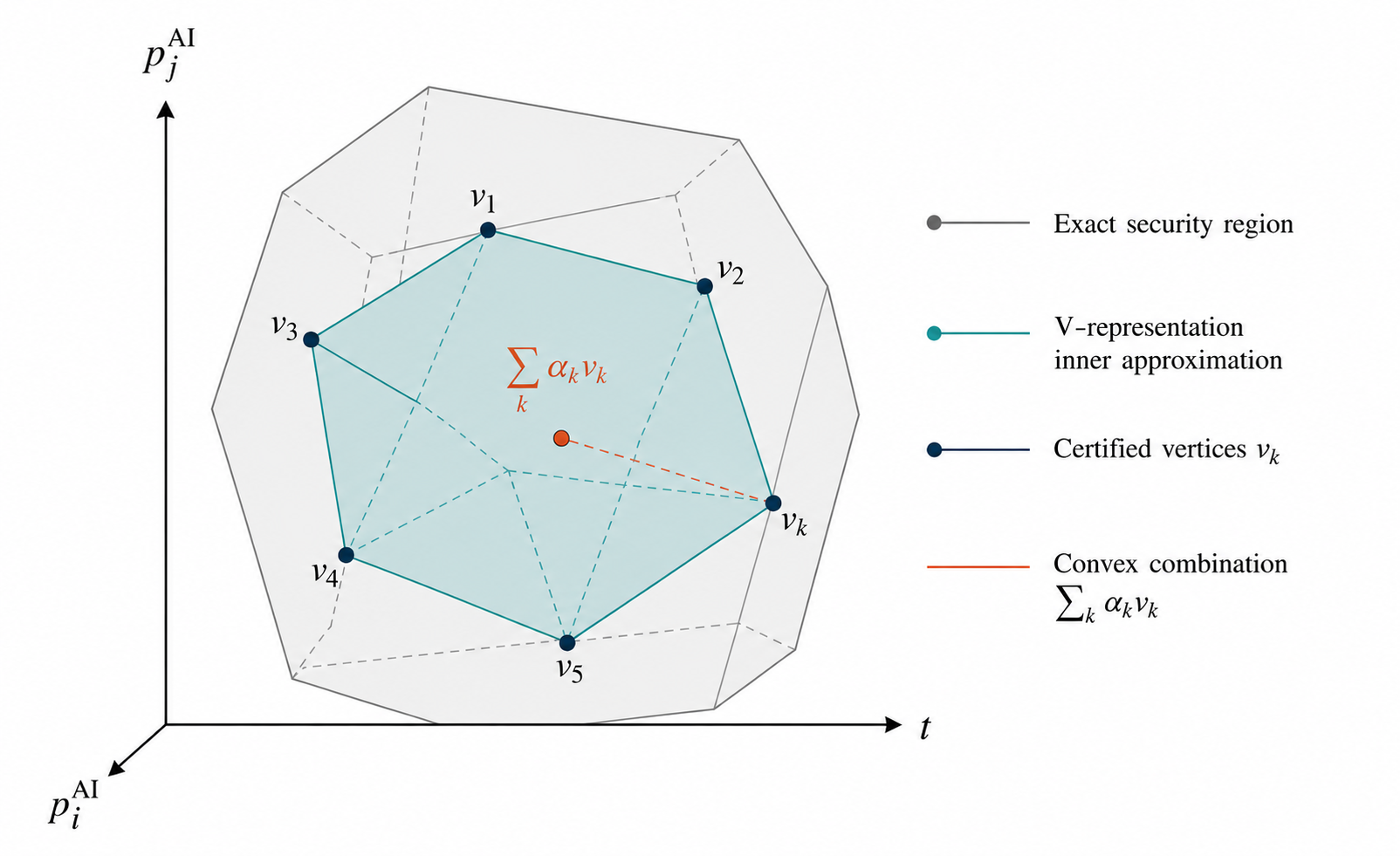}
    \caption{
    %Illustration of the V-representation security-region. 
    V-representation inner approximation of the projected security region.
    %The gray polytope denotes a three-dimensional projection of the exact high-dimensional security region, while the blue polytope denotes the inner approximation formed by certified vertices. The red point represents an admissible AIDC trajectory selected as a convex combination of the transmitted vertices.
    }
    \label{fig:security-region-vrep}
\end{figure}

% Let $\bm{y}^{g}$ collect grid-side continuous variables, including thermal generation, renewable generation, reserve, bus voltage angles, branch flows, storage charging and discharging, and storage state of charge. 
The exact AIDC security region is % the projection
\begin{equation}
    \mcP =
    \left\{
    \bm{x}^{\ai}\in\mathbb{R}_+^{|\mcI|T}
    \middle|
    \exists \bm{y}^{g} \ \mathrm{s.t.}\
    (\bm{x}^{\ai},\bm{y}^{g})\in\mcF^{g}
    \right\},
    \label{eq:step1-projection}
\end{equation}
% \vspace{-0.5cm}
\begin{equation}
    \bm{x}^{\ai}
    =
    \mathrm{col}\{p_{i,t}^{\ai}: i\in\mcI,\ t\in\mcT\}
    \in \mathbb{R}_+^{|\mcI|T}.
    \label{eq:ai-vector}
\end{equation}
where $\mcP$ denotes the projected AIDC security region, 
$\mcT=\{1,\ldots,T\}$ denotes the scheduling periods,
$\mcI$ denotes the set of AIDCs, 
$\bm{x}^{\ai}$ is the full-horizon AIDC power vector, $\bm{y}^g$ denotes grid operating decision variables including thermal generation, renewable generation, reserve, bus voltage angles, branch flows, storage charging and discharging, and storage state of charge, $p_{i,t}^{\ai}$ denotes the active power consumption of AIDC $i$. $\mcF^g$ is the grid feasible set. 
% Directly characterizing $\mcP$ by all of its projected inequalities is computationally expensive and difficult to communicate to the AIDC operator. 
The projected security region $\mcP$ is a high-dimensional polytope whose exact representation cannot be directly shared with the AIDC operator due to privacy concerns.
In this paper, an inner approximation is constructed in V-representation form:
\begin{equation}
    \widehat{\mcP}_{\mathrm{in}}
    =
    \left\{
    \bm{x}^{\ai} =
    \sum_{k\in\mcK} \alpha_k \bm{v}_k
    \middle|
    \sum_{k\in\mcK}\alpha_k=1,\ 
    \alpha_k\ge 0
    \right\},
    \label{eq:step1-vrep}
\end{equation}
where $\widehat{\mcP}_{\mathrm{in}}$ is the inner approximation of the security region, $\bm{v}_k$ is the $k$th certified AIDC power trajectory, and $\alpha_k$ is the convex combination weight. 
The key idea is to identify sufficiently many certified vertices in the high-dimensional feasible set and then construct their convex hull as a communicable inner approximation, as illustrated in Fig.~\ref{fig:security-region-vrep}. Since $\widehat{\mcP}_{\mathrm{in}}\subseteq\mcP$, any trajectory selected from \eqref{eq:step1-vrep} is guaranteed to be secure for grid operation.

Therefore, the security-region interface is decomposed into two technical components: modeling the high-dimensional grid security region and generating representative certified vertices.

\subsection{Grid Feasibility Model}

For each bus $n\in\mcN$, line $\ell\in\mcL$, thermal generator $g\in\mcG$, renewable generator $r\in\mcR$, and storage device $s\in\mcS$, the grid feasible set $\mcF^g$ is defined by
\begingroup
\setlength{\abovedisplayskip}{3pt}
\setlength{\belowdisplayskip}{3pt}
\setlength{\abovedisplayshortskip}{3pt}
\setlength{\belowdisplayshortskip}{3pt}
\setlength{\jot}{1pt}
\begin{align}
    &f_{\ell,t}
    =
    b_{\ell}(\theta_{m,t}-\theta_{n,t}-\phi_{\ell}),
    && \forall \ell=(m,n),t, \label{eq:step1-dcflow}\\
    &-\overline F_{\ell}\le f_{\ell,t}\le \overline F_{\ell},
    && \forall \ell,t, \label{eq:step1-line}\\
    &\underline P_g \le p_{g,t}^{g} \le \overline P_g,
    && \forall g,t, \label{eq:step1-gen-capacity}\\
    &p_{g,t}^{g}+r_{g,t}^{+}\le \overline P_g,\quad
    p_{g,t}^{g}-r_{g,t}^{-}\ge \underline P_g,
    && \forall g,t, \label{eq:step1-gen-reserve}\\   
    &\sum_{g\in\mcG}r_{g,t}^{+}\ge R_t^{+},\quad
    \sum_{g\in\mcG}r_{g,t}^{-}\ge R_t^{-},
    && \forall t, \label{eq:step1-reserve}\\
    &-R_g^{-}\le p_{g,t}^{g}-p_{g,t-1}^{g}\le R_g^{+},
    && \forall g,\ t\in\mcT\setminus\{1\}, \label{eq:step1-ramp}\\
    &\underline P_{r,t}^{r}\le p_{r,t}^{r}\le \overline P_{r,t}^{r},
    && \forall r,t, \label{eq:step1-res}\\
    &0\le p_{s,t}^{\mathrm{ch}}\le \overline P_s^{\mathrm{ch}},\quad
    0\le p_{s,t}^{\mathrm{dis}}\le \overline P_s^{\mathrm{dis}},
    && \forall s,t, \label{eq:step1-storage-power}\\
    &\underline E_s\le e_{s,t}\le \overline E_s,
    && \forall s,t, \label{eq:step1-storage-energy}
 \end{align}
 \begin{align}
    &e_{s,t}
    =
    e_{s,t-1}
    + \eta_s^{\mathrm{ch}}p_{s,t}^{\mathrm{ch}}\dt
    - \frac{p_{s,t}^{\mathrm{dis}}\dt}{\eta_s^{\mathrm{dis}}},
    && \forall s,t, \label{eq:step1-storage}\\
    &e_{s,T}=e_{s,0},
    && \forall s. \label{eq:step1-storage-terminal}
\end{align}
\endgroup
\begingroup
\setlength{\abovedisplayskip}{3pt}
\setlength{\belowdisplayskip}{3pt}
\setlength{\abovedisplayshortskip}{3pt}
\setlength{\belowdisplayshortskip}{3pt}
\begin{align}
    &\sum_{g\in\mcG_n}p_{g,t}^{g}
    + \sum_{r\in\mcR_n}p_{r,t}^{r}
    + \sum_{s\in\mcS_n}(p_{s,t}^{\mathrm{dis}}-p_{s,t}^{\mathrm{ch}})
    - d_{n,t}
    - \sum_{i\in\mcI_n}p_{i,t}^{\ai}
    \notag
    \\[-0.2ex]
    &\quad =
    \sum_{\ell\in\delta_n^+}f_{\ell,t}
    -
    \sum_{\ell\in\delta_n^-}f_{\ell,t},
    \quad \forall n,t.
    \label{eq:step1-balance}
\end{align}
\endgroup
where 
% $\mcG_n$, $\mcR_n$, $\mcS_n$, and $\mcI_n$ are the sets of thermal generators, renewable generators, storage devices, and AIDCs connected to bus $n$, respectively. 
$p_{g,t}^{g}$, $p_{r,t}^{r}$, and $p_{i,t}^{\ai}$ denote thermal generation, renewable generation, and AIDC power consumption; $p_{s,t}^{\mathrm{ch}}$, $p_{s,t}^{\mathrm{dis}}$, and $e_{s,t}$ denote storage charging, discharging, and energy state; $d_{n,t}$ is the conventional load; $f_{\ell,t}$ is the branch flow; and $\theta_{n,t}$ is the bus voltage angle. The parameters $\overline F_\ell$, $\underline P_g$, $\overline P_g$, $R_t^{+}$, and $R_t^{-}$ denote branch limits, generator output limits, and reserve requirements. Renewable output is bounded by $\underline P_{r,t}^{r}$ and $\overline P_{r,t}^{r}$. The terminal condition prevents the security region from being enlarged by depleting storage at the end of the horizon.

Equations \eqref{eq:step1-dcflow}--\eqref{eq:step1-line} describe DC branch flows and line limits. Equations \eqref{eq:step1-gen-capacity}--\eqref{eq:step1-ramp} impose thermal output, reserve capability, reserve requirements, and ramping constraints. Equation \eqref{eq:step1-res} bounds renewable dispatch. Equations \eqref{eq:step1-storage-power}--\eqref{eq:step1-storage-terminal} describe storage power limits, energy limits, intertemporal dynamics, and terminal restoration. Finally, \eqref{eq:step1-balance} enforces nodal power balance including conventional load and AIDC consumption.

\subsection{Vertex Generation}

The vertices in \eqref{eq:step1-vrep} are generated by solving a sequence of linear programs over $\mcF^g$. For a direction matrix $\bm{W}_q\in\mathbb{R}^{|\mcI|\times T}$, the $q$th vertex is obtained from
\begin{equation}
    \max_{(\bm{x}^{\ai},\bm{y}^{g})\in\mcF^g}
    \sum_{i\in\mcI}\sum_{t\in\mcT}
    W_{q,i,t}p_{i,t}^{\ai}.
    \label{eq:step1-vertex-lp}
\end{equation}
where $\bm{W}_q$ is the direction matrix used to expose one boundary point of the security region. For example, maximizing the total AIDC power over all sites and periods corresponds to an all-one direction matrix, i.e., $W_{q,i,t}=1$ for all $i$ and $t$. Several classes of directions can be used, including total AIDC power maximization/minimization, training-only and inference-only directions, per-AIDC directions, time-window directions, and deterministic random directions. In general, a larger vertex set yields a tighter V-representation inner approximation. 
These certified vertices are passed to the AIDC operator for AI workload allocation problem.

\section{Phase II: AI Workload Allocation}
\label{sec:step2}

Given the grid security region, the AIDC operator determines a day-ahead workload schedule that balances computing performance and grid security. The scheduling problem allocates heterogeneous AI workloads among AIDCs and generates an AIDC power trajectory that remains inside the certified region provided by the grid operator.

Two types of workload flexibility are exploited in the proposed AIDC side model. Spatial flexibility is captured by routing inference workload among different Inference AIDCs. Temporal flexibility is not achieved through workload postponement. 
%Instead, it is enabled by DVFS, which adjusts the power consumption and computational throughput within each scheduling interval. 
Instead, it is enabled by DVFS, which adjusts computational throughput to regulate power consumption within each scheduling interval. 
The entire workload allocation problem is formulated as an optimization model and solved to obtain the coordinated AIDC operating schedule. 
Based on the results, checkpoint alert information, including candidate timing windows and drop magnitudes, is identified and communicated to the grid operator for Phase III.

\subsection{AIDC Workload Classification}

The AIDC operator manages two types of facilities with distinct power consumption characteristics and workload demand~\cite{force2025characteristics}.
\begin{itemize}
    \item Training AIDCs are large capacity sites dominated by pre-training and fine-tuning workloads. In practice, post-training may include continued pre-training, supervised fine-tuning (SFT), preference optimization, and distillation. For simplification, only fine-tuning workloads are explicitly modeled in this paper. Their power consumption can be adjusted through DVFS.
    %, while pre-training jobs may also induce checkpoint-related short-term power drops.
    \item Inference AIDCs are smaller sites serving real-time user inference (RT) and latency-tolerant inference (LT) demand. RT demand must be processed in the current period, whereas LT demand can be operated under DVFS modes~\cite{colangelo2026ai}. 
    %Inference workload can also be routed across Inference AIDCs with an explicit remote-processing penalty.
\end{itemize}

\subsection{Grid Security Constraint}

Let $\bm{v}_k$ be the $k$th certified AIDC power trajectory vector provided by the power grid in Phase I.
% , and let $v_{k,i,t}$ be its component for AIDC $i$ at period $t$. 
A convex combination is selected as follows:
\begin{alignat}{2}
    p_{i,t}^{\ai} &=
    \sum_{k\in\mcK}\alpha_k v_{k,i,t},
    && \forall i\in\mcI,\ t\in\mcT,
    \label{eq:step2-vrep-power}\\
    \sum_{k\in\mcK}\alpha_k &= 1,\quad
    \alpha_k\ge 0.
    \label{eq:step2-alpha}
\end{alignat}
where $p_{i,t}^{\ai}$ is the scheduled power of AIDC $i$, $v_{k,i,t}$ is the corresponding component of the $k$th security region vertex, and $\alpha_k$ is the convex combination weight.
%Each AIDC has a site-level capacity limit:
%\begin{equation}
%    0\le p_{i,t}^{\ai}\le \overline P_i^{\ai}.
%    \label{eq:step2-capacity}
%\end{equation}
%where $\overline P_i^{\ai}$ is the rated power capacity of AIDC $i$. 
Equations \eqref{eq:step2-vrep-power}--\eqref{eq:step2-alpha} ensure that the workload schedule remains inside the grid security region.

\subsection{DVFS Mode Representation}

DVFS provides significant operational flexibility for AIDC workloads by enabling controllable trade-offs between computational throughput and power consumption. The measured sweep data in~\cite{colangelo2026ai} show a nonlinear throughput--power relation, which is represented by several operating points and a piecewise-linear approximation in this paper. 
Each mode $m$ corresponds to an operating point $(\rho_m,\eta_m)$, where $\rho_m$ and $\eta_m$ denote the normalized power and throughput, respectively. 
% The throughput--power characteristic is represented by a piecewise-linear curve connecting adjacent operating points. 
% Therefore, any feasible operating point must lie on one of these line segments and can be expressed as a linear combination of two adjacent operating points.

%is an operating point $(\rho_m,\eta_m)$, where $\rho_m$ is the normalized power ratio and $\eta_m$ is the normalized throughput. A linear combination of two adjacent operating points corresponds to one segment of the piecewise-linear throughput--power relation.
\begin{figure}[!t]
    \centering
    \includegraphics[width=\linewidth]{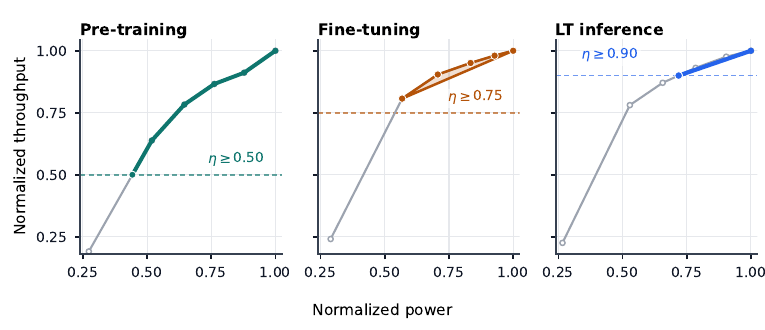}
    \caption{
    DVFS throughput--power relation and operation regions for different workloads
    % approximations and admissible operating regions for pre-training, fine-tuning, and LT inference workloads.
    }
    \label{fig:dvfs-pwl}
\end{figure}

The throughput--power relationships of different AIDC workloads for a single GPU are illustrated in Fig.~\ref{fig:dvfs-pwl}. For pre-training workloads, all GPUs within a cluster must remain synchronized. Therefore, the aggregate operating point must lie on the corresponding piecewise-linear throughput--power curve. 
In contrast, for fine-tuning workloads, each user request activates only a small subset of GPUs relative to the cluster scale. At the cluster level, the aggregate operating point can be represented as a weighted combination of multiple task requests.
% and is not restricted to a single operating segment. 
The feasible operating region can be characterized by the convex hull of multiple operating points. 
The admissible throughput reductions for these workloads are limited to  $50\%$ and $25\%$, respectively.
% For LT-inference workloads, the admissible throughput reduction is limited to $10\%$. For simplicity, the corresponding throughput--power relationship is approximated as linear.
For LT-inference workloads, the admissible throughput reduction is limited to $10\%$. The operating range is relatively narrow, and the corresponding throughput--power relationship is approximated as linear for modeling simplicity.
The sets $\mcM^{\pt}$, $\mcM^{\ft}$, and $\mcM^{\lt}$ collect the retained DVFS modes for pre-training, fine-tuning, and LT inference, respectively. 
Let $\mathcal Q^{\pt}$ denote the set of adjacent pre-training DVFS segments, and let $\mathcal Q_m^{\pt}$ denote the subset of segments incident to mode $m$. The interpolation weights for training cluster $c$ are $\lambda_{c,t,m}^{\pt}$ and $\lambda_{c,t,m}^{\ft}$, and the binary segment-selection variable for pre-training is $\delta_{c,t,q}^{\pt}$.
The pre-training DVFS point is constrained by
\begin{align}
    \sum_{m\in\mcM^{\pt}}\lambda_{c,t,m}^{\pt}
    &=1,
    && \forall c\in\mcC_i^{\pt},\ t\in\mcT, \notag\\
    \sum_{q\in\mathcal Q^{\pt}}\delta_{c,t,q}^{\pt}
    &=1,
    && \forall c\in\mcC_i^{\pt},\ t\in\mcT, \notag\\
    0\le \lambda_{c,t,m}^{\pt}
    &\le
    \sum_{q\in\mathcal Q_m^{\pt}}\delta_{c,t,q}^{\pt},
    && \forall c\in\mcC_i^{\pt},\ t\in\mcT.
    % \ m\in\mcM^{\pt}.
    \label{eq:step2-pretrain-dvfs}
\end{align}
where 
% $\mcM^{\pt}$, $\mcM^{\ft}$, and $\mcM^{\lt}$ are the DVFS mode sets for pre-training, fine-tuning, and LT-inference workloads, respectively,
$\mcQ^{\pt}$ is the set of adjacent pre-training DVFS segments, $\mcQ_m^{\pt}$ is the subset of segments incident to mode $m$, 
$\lambda_{c,t,m}^{\pt}$ and $\lambda_{c,t,m}^{\ft}$ are interpolation weights, 
$\delta_{c,t,q}^{\pt}$ is a binary segment selection variable.
% Thus, all nonincident $\lambda$ variables are forced to zero, and the selected point lies on one piecewise-linear segment.
For fine-tuning, only the convex combination condition is imposed:
\begin{align}
    \sum_{m\in\mcM^{\ft}}\lambda_{c,t,m}^{\ft}
    &=1,
    && \forall c\in\mcC_i^{\ft},\ t\in\mcT, \notag\\
    \lambda_{c,t,m}^{\ft}
    &\ge 0,
    && \forall c\in\mcC_i^{\ft},\ t\in\mcT,\ m\in\mcM^{\ft}.
    \label{eq:step2-finetune-dvfs}
\end{align}

\subsection{Training AIDC Workload}

Let $\mcC_i^{\pt}$ and $\mcC_i^{\ft}$ denote the pre-training and fine-tuning cluster sets in training AIDC $i$. For cluster $c$, let $\overline P_c$ be its nominal power. Given the DVFS representation in \eqref{eq:step2-pretrain-dvfs}--\eqref{eq:step2-finetune-dvfs}, the training power at site $i$ is
\begin{align}
    p_{i,t}^{\tr}
    =
    \sum_{c\in\mcC_i^{\pt}}\overline P_c
    \sum_{m\in\mcM^{\pt}}\rho_m^{\pt}\lambda_{c,t,m}^{\pt} 
    % \notag\\
    +
    \sum_{c\in\mcC_i^{\ft}}\overline P_c
    \sum_{m\in\mcM^{\ft}}\rho_m^{\ft}\lambda_{c,t,m}^{\ft}.
    \label{eq:step2-training-power}
\end{align}
where $p_{i,t}^{\tr}$ is the total power of training AIDC $i$.
% , and $\rho_m^{\pt}$ and $\rho_m^{\ft}$ are normalized power ratios. For a Training AIDC, $p_{i,t}^{\ai}=p_{i,t}^{\tr}$.

\subsection{Inference AIDC Workload}

Let $\mcI^{\infm}$ denote the set of Inference AIDCs. Each inference AIDC has its own RT and LT demand, denoted by $d_{o,t}^{\rt}$ and $d_{o,t}^{\lt}$.
%, where $o$ is the workload origin.
RT demand is processed without DVFS degradation, whereas LT demand is represented by the two LT operating points in $\mcM^{\lt}$, shown in Fig.~\ref{fig:dvfs-pwl}. 
% The decision variable $y_{o,j,t}^{\rt}$ is the amount of RT workload generated at origin $o$ and processed at inference site $j$. 
% The variable $u_{o,j,t,m}^{\lt}$ is the LT workload generated at $o$, processed at $j$, and assigned to LT operating point $m$. 
Workload demand satisfaction is enforced by
\begin{align}
    \sum_{j\in\mcI^{\infm}}y_{o,j,t}^{\rt}
    &= d_{o,t}^{\rt},
    && \forall o\in\mcI^{\infm},\ t\in\mcT,
    \label{eq:step2-rt-demand}\\
    \sum_{j\in\mcI^{\infm}}\sum_{m\in\mcM^{\lt}}
    u_{o,j,t,m}^{\lt}
    &= d_{o,t}^{\lt},
    && \forall o\in\mcI^{\infm},\ t\in\mcT.
    \label{eq:step2-lt-demand}
\end{align}
where $d_{o,t}^{\rt}$ and $d_{o,t}^{\lt}$ are the RT and LT demands of origin inference site $o$, 
$y_{o,j,t}^{\rt}$ denotes the RT workload routed from AIDC $o$ to AIDC $j$, 
and $u_{o,j,t,m}^{\lt}$ is the LT workload routed from AIDC $o$ to AIDC $j$ under DVFS mode $m$.
Thus, each Inference AIDC can process its own workload or route it to another Inference AIDC. Remote processing is allowed but penalized in the objective.

Let $a_j^{\rt}$ and $a_j^{\lt}$ be the workload-to-power conversion coefficients of Inference AIDC $j$. The power consumption of inference AIDC is
\begin{align}
    p_{j,t}^{\infm}
    =
    a_j^{\rt}\sum_{o\in\mcI^{\infm}}y_{o,j,t}^{\rt}
    +
    a_j^{\lt}\sum_{o\in\mcI^{\infm}}\sum_{m\in\mcM^{\lt}}
    \rho_m^{\lt}u_{o,j,t,m}^{\lt}.
    \label{eq:step2-inference-power}
\end{align}
where $p_{j,t}^{\infm}$ is the power of Inference AIDC $j$, and $a_j^{\rt}$ and $a_j^{\lt}$ are workload-to-power conversion coefficients. 
% For an Inference AIDC, $p_{j,t}^{\ai}=p_{j,t}^{\infm}$.

\subsection{Objective Function}
The AI service utility is quantified by the total GPU throughput delivered across all workloads. The objective maximizes the total service utility while minimizing cross-site processing penalties for inference workload.
\begin{align}
    \max\quad
    & U^{\tr}+U^{\lt}
    - C^{\mathrm{rem}},
    \label{eq:step2-objective}
\end{align}
where $C^{\mathrm{rem}}$ is the remote-processing penalty. The training AIDC service utility is
\begin{align}
    U^{\tr}
    =
    &\sum_{i\in\mcI^{\tr}}\sum_{c\in\mcC_i^{\pt}}\sum_{t,m}
    \overline P_c\eta_m^{\pt}\lambda_{c,t,m}^{\pt}
    \notag\\
    &+
    \sum_{i\in\mcI^{\tr}}\sum_{c\in\mcC_i^{\ft}}\sum_{t,m}
    \overline P_c\eta_m^{\ft}\lambda_{c,t,m}^{\ft},
    \label{eq:step2-training-utility}
\end{align}
where $\eta_m^{\pt}$ and $\eta_m^{\ft}$ are normalized throughputs of pre-training and fine-tuning DVFS modes. The LT-inference utility is
\begin{equation}
    U^{\lt}
    =
    \sum_{o,j,t,m}
    \eta_m^{\lt}u_{o,j,t,m}^{\lt}.
    \label{eq:step2-lt-utility}
\end{equation}
where $\eta_m^{\lt}$ is the LT-inference normalized throughput. The remote-processing penalty is
\begin{equation}
    C^{\mathrm{rem}}
    =
    \sum_{\substack{o,j,t\\o\ne j}}
    c_{\rt}^{\mathrm{rem}}y_{o,j,t}^{\rt}
    +
    \sum_{\substack{o,j,t,m\\o\ne j}}
    c_{\lt}^{\mathrm{rem}}u_{o,j,t,m}^{\lt}.
    \label{eq:step2-remote}
\end{equation}
where $c_{\rt}^{\mathrm{rem}}$ and $c_{\lt}^{\mathrm{rem}}$ are penalties for remotely processing RT and LT workloads, respectively.

\subsection{Checkpoint Candidate Window}

During large-scale pre-training, checkpoint periodically saves model states and optimizer states to storage for failure recovery. This operation can pause computation on a large number of GPUs, causing a short but deep power reduction in the corresponding training AIDC. The subsequent return to full computation may create a steep load recovery, which is operationally relevant for grid scheduling.

Checkpoint uncertainty is not optimized directly in the workload allocation model. After the workload schedule is determined, the admissible timing window of each checkpoint enabled pre-training cluster is identified. For a checkpoint event $e$, the input parameters specify a baseline period $\tau_e^0$, a timing tolerance, and a retained-power ratio $\kappa_e$. The admissible candidate set is
\begin{equation}
    \mcT_e^{\mathrm{cp}}
    =
    \{\tau: |\tau-\tau_e^0|\le \Delta_e^{\mathrm{tol}}\}.
    \label{eq:step2-cp-window}
\end{equation}
where $\mcT_e^{\mathrm{cp}}$ is the checkpoint candidate window, $\tau_e^0$ is the baseline checkpoint period, and $\Delta_e^{\mathrm{tol}}$ is the timing tolerance.
For candidate $\tau\in\mcT_e^{\mathrm{cp}}$, the checkpoint-induced power reduction is
\begin{equation}
    d_{e,\tau}^{\mathrm{cp}}
    =
    (1-\kappa_e)P_{c,\tau}^{\pt},
    \label{eq:step2-cp-drop}
\end{equation}
where $d_{e,\tau}^{\mathrm{cp}}$ is the checkpoint power drop, 
% $\kappa_e$ is the retained-power ratio, 
and $P_{c,\tau}^{\pt}$ is the scheduled power of the corresponding pre-training cluster $c$ in a Training AIDC. The planned AIDC power trajectory and the checkpoint candidate windows of the Training AIDCs are then communicated to the grid-side robust OPF.

\vspace{-1ex}
\section{Phase III: Checkpoint-Aware Two-Stage Robust OPF}
\label{sec:step3}

After the AIDC power trajectory and checkpoint alert windows are provided, the grid operator performs day-ahead OPF. 
The schedule must remain feasible under renewable generation deviations and the uncertain timing of checkpoint-induced AIDC power drops, leading to a checkpoint-aware two-stage robust OPF. 
% The checkpoint information is not modeled as an unstructured load forecast uncertainty. 
% The admissible checkpoint time windows and load reduction magnitudes are explicitly provided by the AIDC operator and incorporated into the robust uncertainty set.

Since checkpoint data storage typically lasts for 15--30 minutes, the scheduling horizon is discretized at a 15-min resolution, resulting in 96 periods for 24-hour day-ahead operation problem. 
This resolution allows checkpoint-induced load reductions to be represented without averaging out their operational impact. Checkpoint candidates are modeled through a discrete timing uncertainty set.
% , while transmission line flows are represented using the DC power flow model. 
% The resulting robust problem can be solved by C\&CG. 
% The first stage determines generation set-points, renewable set-points, and reserve schedules. The second stage models real-time redispatch, storage operation, renewable curtailment, load shedding, and network feasibility after uncertainty realization.
\vspace{-1ex}
\subsection{Checkpoint-aware AIDC Load}

Let $P_{n,t}^{d,0}$ be the conventional load at bus $n$. The planned AIDC power trajectory is mapped to the corresponding buses:
\begin{equation}
    P_{n,t}^{d}
    =
    P_{n,t}^{d,0}
    +
    \sum_{i\in\mcI_n}p_{i,t}^{\ai}.
    \label{eq:step3-load}
\end{equation}
where $P_{n,t}^{d}$ is the total nodal load used in OPF, and $\mcI_n$ is the set of AIDCs connected to bus $n$.
Checkpoint candidates are indexed by $k\in\mcK^{\mathrm{cp}}$. Candidate $k$ is associated with bus $b_k$, time $\tau_k$, and drop magnitude $d_k^{\mathrm{cp}}$. The binary variable $\chi_k$ selects whether candidate $k$ occurs. 
Because checkpoint reduces training data center consumption, it is modeled as a positive nodal injection:
\begin{equation}
    P_{n,t}^{\mathrm{cp}}(\bm{\chi})
    =
    \sum_{k\in\mcK^{\mathrm{cp}}: b_k=n, \tau_k=t}
    d_k^{\mathrm{cp}}\chi_k .
    \label{eq:step3-cp-inj}
\end{equation}
where $P_{n,t}^{\mathrm{cp}}(\bm{\chi})$ is the checkpoint-induced load reduction at bus $n$ and time $t$.

\subsection{Uncertainty Set}

Renewable uncertainty is represented by upward and downward binary deviations. Let $\widehat P_{r,t}^{\mathrm{ren}}$ be the forecast renewable availability and $\Delta_{r,t}^{\mathrm{ren}}$ be the maximum deviation. The realized renewable availability is
\begin{equation}
    P_{r,t}^{\mathrm{av}}(\bm{z})
    =
    \widehat P_{r,t}^{\mathrm{ren}}
    - \Delta_{r,t}^{\mathrm{ren}}z_{r,t}^{-}
    + \Delta_{r,t}^{\mathrm{ren}}z_{r,t}^{+}.
    \label{eq:step3-ren-avail}
\end{equation}
where $P_{r,t}^{\mathrm{av}}(\bm{z})$ is the realized renewable availability, $\widehat P_{r,t}^{\mathrm{ren}}$ is the forecast availability, $\Delta_{r,t}^{\mathrm{ren}}$ is the deviation magnitude, and $z_{r,t}^{-}$ and $z_{r,t}^{+}$ are binary variables for downward and upward deviations.
The renewable budget set is
\begin{align}
    z_{r,t}^{-}+z_{r,t}^{+} &\le 1,
    && \forall r\in\mcR,\ t\in\mcT,\\
    \sum_{r,t}(z_{r,t}^{-}+z_{r,t}^{+}) &\le \Gamma^{\mathrm{ren}}.
    \label{eq:step3-ren-budget}
\end{align}
where $\Gamma^{\mathrm{ren}}$ is the renewable uncertainty budget.
For each checkpoint event $e$, exactly one candidate timing is selected:
\begin{align}
    \sum_{k\in\mcK_e^{\mathrm{cp}}}\chi_k=1,
    && \forall e\in\mcE^{\mathrm{cp}}.
    % \chi_k\in\{0,1\}, && \forall k\in\mcK^{\mathrm{cp}}.
    \label{eq:step3-cp-budget}
\end{align}
where $\mcE^{\mathrm{cp}}$ is the set of checkpoint events and $\mcK_e^{\mathrm{cp}}$ is the candidate set associated with checkpoint event $e$.
The complete uncertainty set is
\begin{equation}
    \mcU =
    \left\{
    (\bm{z}^{-},\bm{z}^{+},\bm{\chi})
    \middle|
    \eqref{eq:step3-ren-budget},\eqref{eq:step3-cp-budget}
    \right\}.
    \label{eq:step3-uncertainty}
\end{equation}
where $\mcU$ is the joint uncertainty set of renewable deviations and checkpoint timings.

\subsection{First-Stage Problem}

The first-stage decision vector and feasible set are compactly defined as
\begin{equation}
    \bm{x}=\{\bm{p}^{g},\bm{p}^{\mathrm{ren,DA}},\bm{r}^{+},\bm{r}^{-}\},
    \quad
    \mcX =
    \{\bm{x}\mid \eqref{eq:step1-dcflow}\text{--}\eqref{eq:step1-balance}\}.
    \label{eq:step3-first-stage-set}
\end{equation}
where $p_{g,t}^{g}$ is the thermal generation set point, $p_{r,t}^{\mathrm{ren,DA}}$ is the renewable set point, and $r_{g,t}^{+}$ and $r_{g,t}^{-}$ are upward and downward reserves. The feasible set $\mcX$ reuses the grid feasibility constraints defined in Phase I.

The first-stage objective function is
\begin{align}
    C^{\mathrm{DA}}(\bm{x})
    &=
    \sum_{t\in\mcT}
    \Big[
    \dt\sum_g c_g^{e}p_{g,t}^{g}
    \notag\\
    &\quad
%    \notag\\
%    &\quad
%    + \dt\sum_g c_g^{+}r_{g,t}^{+}
%    + \dt\sum_g c_g^{-}r_{g,t}^{-}
%    \notag\\
%    &\quad
    + \dt \cdot c^{\mathrm{sp}}
    \sum_r(\widehat P_{r,t}^{\mathrm{ren}}
    -p_{r,t}^{\mathrm{ren,DA}})
    \Big].
    \label{eq:step3-da-cost}
\end{align}
where $C^{\mathrm{DA}}$ is the day-ahead cost, $c_g^e$ is the thermal energy cost, and $c^{\mathrm{sp}}$ is the renewable curtailment penalty.

\subsection{Second-Stage Recourse}

For a fixed first-stage solution $\bm{x}$ and uncertainty realization $\bm{\xi}\in\mcU$, the second-stage variables are
\begin{equation}
    \bm{y}=
    \{\tilde{\bm{p}}^g,\bm{a}^{+},\bm{a}^{-},
    \tilde{\bm{p}}^{\mathrm{ren}},\bm{p}^{\mathrm{sp}},
    \bm{p}^{\mathrm{shed}},
    \bm{p}^{\mathrm{ch}},\bm{p}^{\mathrm{dis}},\bm{e}\}.
\end{equation}
where $\bm{y}$ is the second-stage recourse vector; $\tilde p_{g,t}^{g}$ is real-time generation; $a_{g,t}^{+}$ and $a_{g,t}^{-}$ are upward and downward adjustments from the day-ahead set-point; $\tilde p_{r,t}^{\mathrm{ren}}$ is renewable dispatch; $p_{r,t}^{\mathrm{sp}}$ is renewable curtailment; $p_{n,t}^{\mathrm{shed}}$ is load shedding; and $\bm{p}^{\mathrm{ch}}$, $\bm{p}^{\mathrm{dis}}$, and $\bm{e}$ denote storage charging, discharging, and energy states.

The redispatch relation and reserve deployment are
\begin{align}
    \tilde p_{g,t}^{g}
    &=
    p_{g,t}^{g}+a_{g,t}^{+}-a_{g,t}^{-},
    && \forall g\in\mcG,\ t\in\mcT,\\
    0\le a_{g,t}^{+}\le r_{g,t}^{+},
    &\quad
    0\le a_{g,t}^{-}\le r_{g,t}^{-},
    && \forall g\in\mcG,\ t\in\mcT,\\
    \underline P_g
    \le \tilde p_{g,t}^{g}
    &\le \overline P_g,
    && \forall g\in\mcG,\ t\in\mcT.
    \label{eq:step3-redispatch}
\end{align}
where $\tilde p_{g,t}^{g}$ is coupled to the day-ahead set-point by reserve deployment variables $a_{g,t}^{+}$ and $a_{g,t}^{-}$.
Real-time generation also satisfies ramping limits. Storage constraints follow the same energy dynamics as in \eqref{eq:step1-storage}, with terminal energy restored to the initial value.

Renewable dispatch and curtailment satisfy
\begin{align}
    & 0\le \tilde p_{r,t}^{\mathrm{ren}}
    \le P_{r,t}^{\mathrm{av}}(\bm{z}),
    && \forall r\in\mcR,\ t\in\mcT,\\
    p_{r,t}^{\mathrm{sp}}
    &=
    P_{r,t}^{\mathrm{av}}(\bm{z})
    -\tilde p_{r,t}^{\mathrm{ren}}
    %,
    % \quad
    %p_{r,t}^{\mathrm{sp}}
    \ge 0,
    && \forall r\in\mcR,\ t\in\mcT.
\end{align}
where $\tilde p_{r,t}^{\mathrm{ren}}$ is real-time renewable dispatch and $p_{r,t}^{\mathrm{sp}}$ is renewable curtailment.
Load shedding is bounded by the net load after checkpoint reduction:
\begin{equation}
    0\le p_{n,t}^{\mathrm{shed}}
    \le P_{n,t}^{d}-P_{n,t}^{\mathrm{cp}}(\bm{\chi}),
    \quad \forall n\in\mcN,\ t\in\mcT.
    \label{eq:step3-shed-bound}
\end{equation}
where $p_{n,t}^{\mathrm{shed}}$ is the load-shedding variable.

Let $C_g$, $C_r$, and $C_s$ be bus incidence matrices for thermal generators, renewable generators, and storage. The real-time nodal injection is
\begin{align}
    \bm{\iota}_t^{\mathrm{RT}}
    =
    &C_g\tilde{\bm{p}}_t^{g}
    +C_r\tilde{\bm{p}}_t^{\mathrm{ren}}
    +C_s(\bm{p}_t^{\mathrm{dis}}-\bm{p}_t^{\mathrm{ch}})
    \notag\\
    &+\bm{P}_t^{\mathrm{cp}}(\bm{\chi})
    -\bm{P}_t^{d}
    +\bm{p}_t^{\mathrm{shed}}.
    \label{eq:step3-injection}
\end{align}
% where $\bm{\iota}_t^{\mathrm{RT}}$ is the real-time net injection vector, $C_g$, $C_r$, and $C_s$ are bus incidence matrices, and 
Power balance and PTDF line-flow constraints are
\begin{equation}
    \bm{1}^{\top}\bm{\iota}_t^{\mathrm{RT}}=0,\quad
    -\overline{\bm{F}}\le H\bm{\iota}_t^{\mathrm{RT}}
    \le \overline{\bm{F}},
    \quad \forall t\in\mcT.
    \label{eq:step3-network}
\end{equation}
where $H$ is the PTDF matrix, and $\overline{\bm F}$ is the vector of branch-flow limits.

The second-stage cost is
\begin{align}
    C^{\mathrm{RT}}(\bm{y};\bm{\xi})
    =
    \sum_{t\in\mcT}
    \Big[
    &\dt\sum_g c_g^{\mathrm{rd}}
    (a_{g,t}^{+}+a_{g,t}^{-})
    \notag\\
    &+\dt\sum_s c_s^{\mathrm{ess}}
    (p_{s,t}^{\mathrm{ch}}+p_{s,t}^{\mathrm{dis}})
    \notag\\
    &+\dt c^{\mathrm{sp}}\sum_r p_{r,t}^{\mathrm{sp}}
    +\dt c^{\mathrm{shed}}\sum_n p_{n,t}^{\mathrm{shed}}
    \Big].
    \label{eq:step3-rt-cost}
\end{align}
where $C^{\mathrm{RT}}$ is the real-time recourse cost, $c_g^{\mathrm{rd}}$ is the redispatch cost, $c_s^{\mathrm{ess}}$ is the storage operation cost, and $c^{\mathrm{shed}}$ is the load-shedding penalty.
\vspace{-2ex}
\subsection{Two-Stage Robust Formulation}

The compact formulation of the entire problem is
\begin{align}
    \min_{\bm{x}\in\mcX}
    \left\{
    C^{\mathrm{DA}}(\bm{x})
    +
    \max_{\bm{\xi}\in\mcU}
    \min_{\bm{y}\in\mcY(\bm{x},\bm{\xi})}
    C^{\mathrm{RT}}(\bm{y};\bm{\xi})
    \right\},
    \label{eq:step3-robust}
\end{align}
% where $\mcY(\bm{x},\bm{\xi})$ is the recourse feasible set under first-stage schedule $\bm{x}$ and uncertainty realization $\bm{\xi}$. The formulation minimizes the day-ahead cost plus the worst-case real-time recourse cost.
Compared with conventional two-stage robust OPF, the proposed formulation introduces checkpoint timing as an additional binary uncertain variable. This extension does not destroy the convexity of the inner recourse problem, because the checkpoint variable only appears in the outer uncertainty realization, whereas the second-stage dispatch remains a linear program once the uncertainty is fixed. Therefore, the problem retains the standard structure required by C\&CG while explicitly capturing checkpoint-induced AIDC load reductions.

\vspace{-1ex}
\section{Case Study}
\label{sec:case}

In this section, the proposed coordination scheme is evaluated on a modified IEEE 14-bus system with multiple AIDCs and renewable generators. To further demonstrate the scalability and practical applicability, additional case studies are conducted on a modified New York ISO system. Detailed parameter settings of both test systems can be found in~\cite{liu_parameters_ieee14_nyiso}.
% ADD GITHUB LINK

The simulations are implemented in Python~3.13.5 and solved by Gurobi~12.0.3 on a computer with an Apple M4 processor and 16GB memory.
\vspace{-2ex}
\subsection{System Description}

The proposed scheme is first tested on the 14-bus system over a 24-hour horizon with 96 intervals, corresponding to a 15-min resolution. The system contains three thermal units with a total capacity of 780 MW, two wind generators at buses 3 and 8 with installed capacities of 250 MW and 200 MW, respectively, and one 50 MW/200 MWh energy storage system at bus 8. The energy storage system is operated within a 0.2--0.8 SOC range, with an initial SOC of 0.5. % The non-AIDC load follows the original IEEE 14-bus load distribution scaled by a daily profile, resulting in a daily energy of 10,194.24 MWh and a peak demand of 449.48 MW.
The peak demand and total daily energy consumption of the conventional load are 449.48 MW and 10,194.24 MWh, respectively.

Three AIDCs are added to the system: one 300 MW training AIDC at bus 9 and two 50 MW inference AIDCs at buses 6 and 13. The PUE is set to 1.15 for all AIDCs. 
The Training AIDC contains four computing clusters that can be used for both pre-training and fine-tuning workloads, each with a GPU-side rated power of 65.22 MW. Since the number of fine-tuning tasks and the associated GPU utilization are significantly smaller than those of pre-training workloads, only pre-training workloads are considered in this study.
 %Checkpoint timing has a one-period tolerance, producing 16 checkpoint events and 48 candidate checkpoint periods over the horizon. 
The LT and RT demands curve of the Inference AIDC are specified as input parameters. The total peak power demand of the AIDC is 354.66 MW.
\vspace{-2ex}
\begin{table}[t]
    \caption{Comparison of Security Region and Feasibility Check}
    \label{tab:step1-method-feasibility}
    \centering
    \scriptsize
    \setlength{\tabcolsep}{3pt}
    \resizebox{\linewidth}{!}{%
    \begin{tabular}{@{}lcrrrr@{}}
        \toprule
        \multirow{2}{*}{\centering Method}
        & \multirow{2}{*}{Feas.}
        & Farkas viol.
        & TR AIDC 
        & TR AIDC
        & Total AIDC \\
        & & (MW) &MIN. (MW) &AVG. (MW) &AVG. (MW) \\
        \midrule
        V-representation& Yes & 0.00 & 234.76 & 292.63 & 326.40 \\
        Support function cuts~\cite{lohne2016equivalence}& No & 21,731.43 & 241.93 & 299.20 & 334.36 \\
        Sampled Farkas cuts~\cite{fischetti2009minimal} & No & 313.47 & 228.57 & 296.43 & 330.83 \\
        \bottomrule
    \end{tabular}
    }
\end{table}
%\vspace{-2ex}
\subsection{Security-Region Calculation}
% The grid-side security-region calculation uses a V-representation inner approximation. 
In Phase I, 165 vertices are generated and their convex hull, representing the security region, is transmitted to the AIDC operator. For comparison, two outer-approximation methods are also tested: a support function cut approximation~\cite{lohne2016equivalence} and a sampled Farkas cut approximation~\cite{fischetti2009minimal}. After solving the workload allocation model in Phase II, the resulting AIDC power trajectory is projected back to the original high-dimensional security region polytope in Phase I for a feasibility check. The result is shown in Table~\ref{tab:step1-method-feasibility}.

The solution of inner V-representation method remains feasible in the original security region polytope. In contrast, both outer-approximation methods become infeasible when mapped back to the original problem. The Farkas violations further indicate that the Farkas cuts method yields a smaller infeasibility level, it still fails to remain within the original feasible region. It is attributed to the fact that Phase II seeks the optimal AIDC workload allocation within the approximated security region. The optimal solution tends to lie near the boundary of the approximated region. For outer approximations, such boundary points may not belong to the original security region, leading to infeasible solutions in original feasible region. 

This effect is also reflected in the planned AIDC power trajectories. The two outer-approximation methods produce slightly more aggressive power schedules. Although the differences in scheduled power are relatively small, they are sufficient to violate the original security constraints.
% Consequently, when these schedules are passed to Phase III, the resulting grid dispatch is more likely to encounter infeasibility or require load shedding to maintain system security.
This confirms why we choose the inner V-representation method for security region calculation: it preserves a direct security guarantee after the workload-allocation problem is solved.

\begin{figure}[t]
    \centering
    \includegraphics[width=\linewidth]{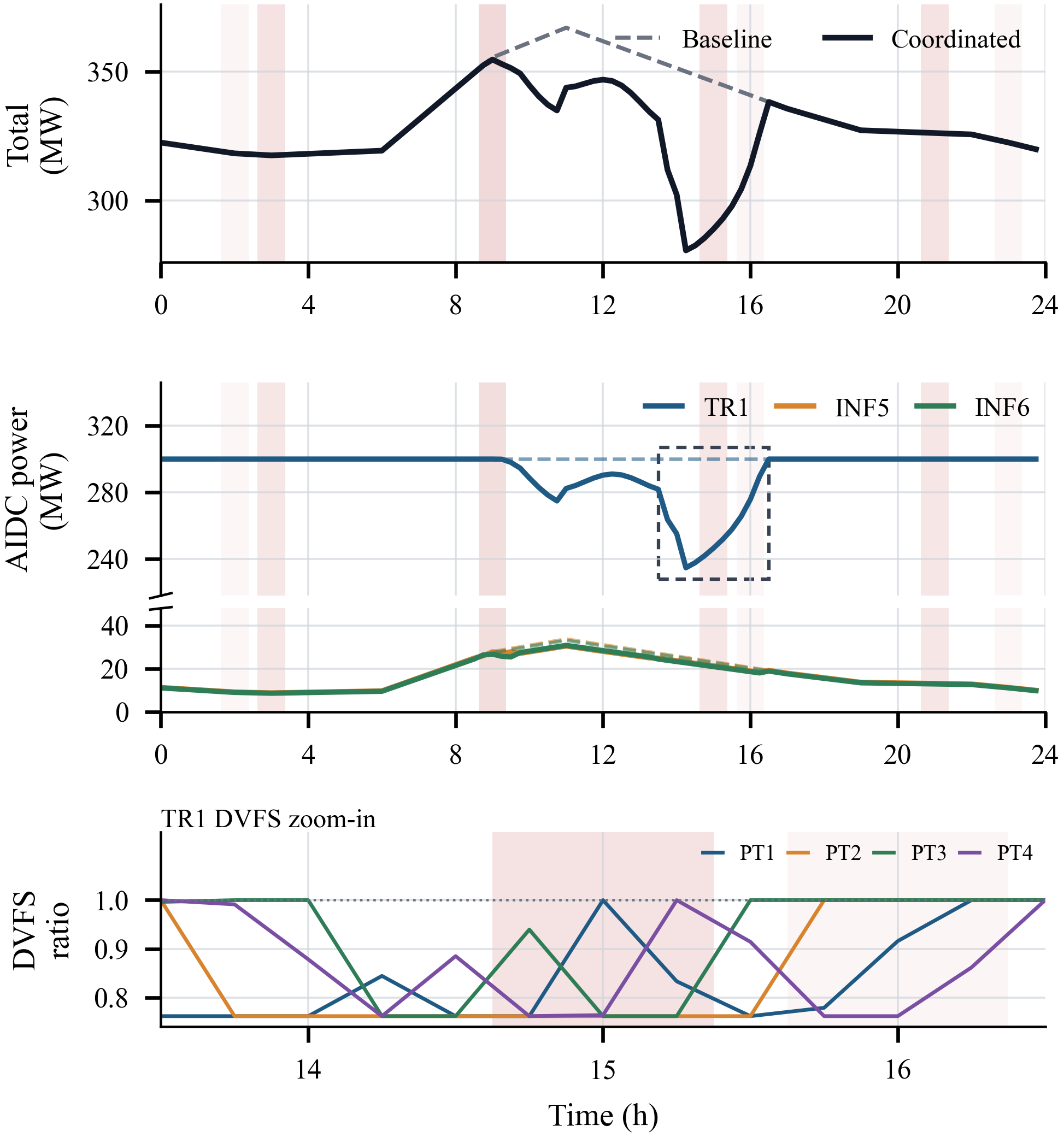}
    \caption{AIDC power trajectories after workload allocation in Phase II. 
   % schedules under the coordinated workload allocation. Dashed curves show the corresponding non-coordinated baselines, shaded regions indicate checkpoint candidate windows, and the inset zooms in on the TR1 DVFS power-ratio adjustment interval.
    }
    \label{fig:step2-power-plan}
    \vspace{-2ex}
\end{figure}
\vspace{-2ex}
\subsection{Workload Allocation Analysis}
In Phase II, the workload allocation results are shown in Fig.~\ref{fig:step2-power-plan}.
To keep the AIDC power trajectory in the security region, the aggregated scheduled AIDC energy is reduced from 8,043.75 MWh to 7,833.69 MWh, corresponding to a reduction of 2.61\%. The Training AIDC operates close to its facility rating for most of the day while still satisfying the transmitted inner approximation.
% In the coordinated solution, the average powers of TR1, INF5, and INF6 are 292.63, 16.98, and 16.80 MW, respectively.

Most of the energy reduction occurs during Hours 10--16, when wind power is at a relatively low level and the grid is under a  tighter supply condition. During this period, all four Training AIDC clusters for pre-training workloads proactively reduce their power consumption through DVFS to alleviate system operating stress. The DVFS ratios of four cluster are also shown in the zoomed-in view of Fig.~\ref{fig:step2-power-plan}. The ratios remain within the range of 0.76--1.00 and are relatively evenly distributed across the four clusters. 
% This balanced allocation strategy avoids excessive throughput degradation of any individual pre-training workload 
The four homogeneous pre-training clusters exhibit similar DVFS ratios in this case
while effectively supporting secure grid operation.

% The allocation result also reveals the different flexibility of workload classes. RT inference and LT inference are processed locally in the tested case to avoid remote-processing penalties. Hence, inference flexibility is represented mainly through DVFS-based power modulation rather than cross-site routing.
%The checkpoint candidate windows identified from the pre-training clusters
The checkpoint candidate windows, provided as AIDC-side operational information, are also shown in Fig.~\ref{fig:step2-power-plan} using light-pink shaded blocks. Across all candidates, the checkpoint induced load drop ranges from 46.86 MW to 61.50 MW, with a mean of 58.59 MW.  The resulting AIDC power trajectories and checkpoint alert information are then transmitted to the grid operator for further grid scheduling.
% \vspace{-2ex}
%The inset of Fig.~\ref{fig:step2-power-plan} zooms in on the main DVFS adjustment interval of TR1; the pre-training cluster power ratios range from 0.762 to 1.000, with a mean of 0.975.

\vspace{-2ex}
\subsection{AI Workload Flexibility Comparison}
\begin{figure}[t]
    \centering
    \includegraphics[width=\linewidth]{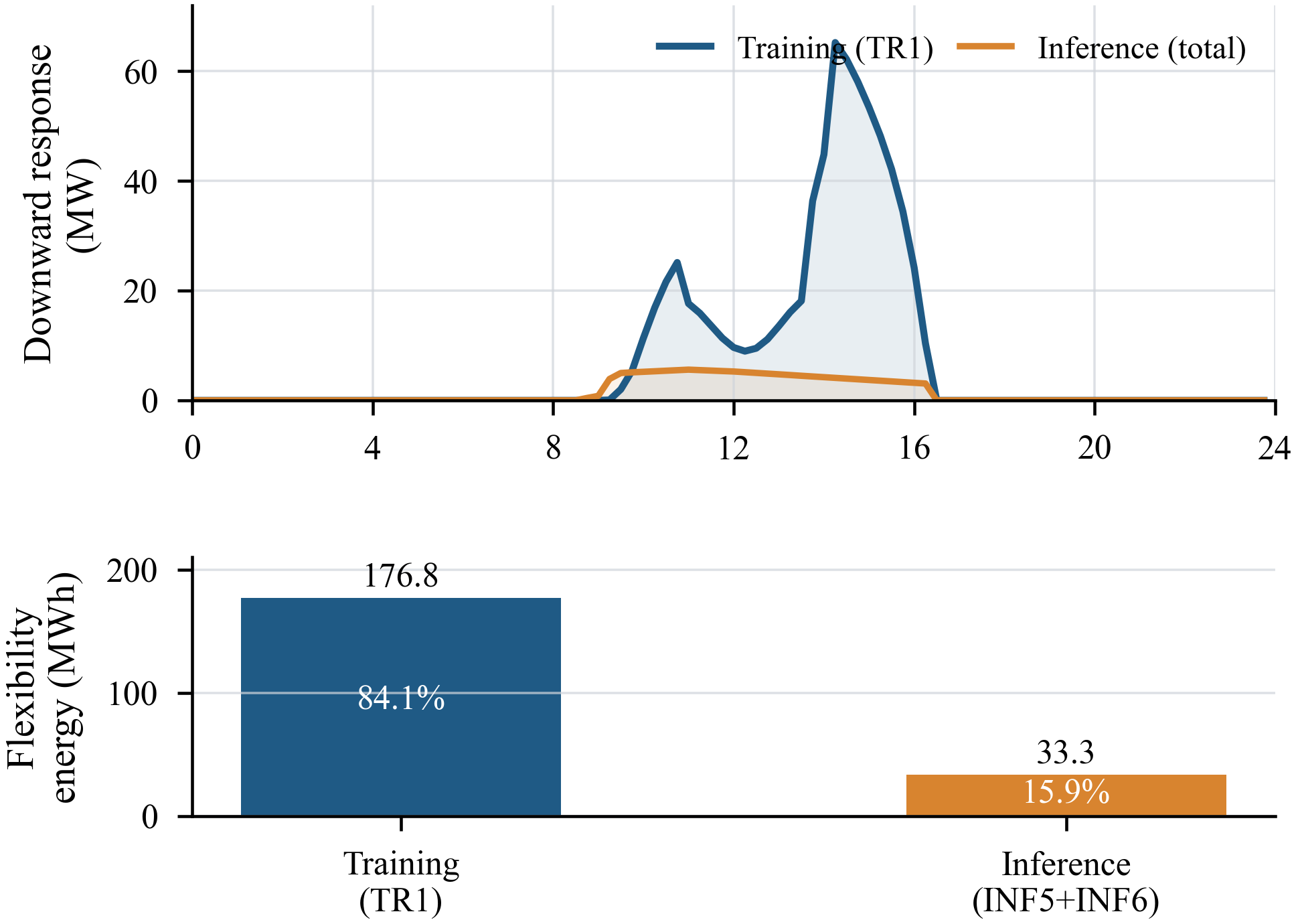}
    \caption{Flexibility of training and inference workload. The upper panel shows the time-varying response, and the lower panel shows the accumulated response energy.}
    \label{fig:phase2-flexibility}
    \vspace{-4ex}
\end{figure}

To evaluate the grid-support potential of different AIDC workloads in Phase II, the flexibility provided by training and inference workloads is compared.
% For each workload class $c$, its realized downward response is defined as $\Delta p_c(t)=\max\{p_c^{\mathrm{ref}}(t)-p_c^{\mathrm{S1}}(t),0\}$. 
Figure~\ref{fig:phase2-flexibility} shows the resulting time profile, and Table~\ref{tab:phase2-flexibility} summarizes the accumulated flexibility. An interval is counted as active when the aggregate downward response exceeds 1 MW.

% Both workload classes are adjusted during the same tight-supply interval, but their response magnitudes are markedly different. 
Training workload supplies 176.83 MWh of downward flexibility, accounting for 84.1\% of the total AIDC response, with a peak response of 65.24 MW. By contrast, the two inference AIDCs together supply 33.34 MWh and a peak response of only 5.59 MW. Although inference is active in a comparable number of intervals (29 versus 28 for Training), its average response intensity is much smaller. 
% Thus, the dominant contribution of Training is driven by response magnitude rather than simply by the number of adjusted time intervals.

This difference follows directly from the workload structures. The 300 MW Training AIDC contains four pre-training clusters whose DVFS profile permits a power ratio down to 0.519, corresponding to a theoretical facility-side downward range of 144.42 MW. In contrast, inference DVFS applies only to the LT workload component and its minimum power ratio is 0.784, while RT demand must be served without degradation. Spatial workload shifting also plays a minor role: only 0.72\% of the LT workload is processed remotely and no RT workload is rerouted. Consequently, Training provides the principal grid-facing flexibility resource in Phase II, whereas the smaller-capacity Inference AIDCs provide only limited supplementary adjustment.

\begin{table}[t]
    \caption{Phase-II Flexibility Activation Under S1}
    \label{tab:phase2-flexibility}
    \centering
    \footnotesize
    \setlength{\tabcolsep}{2.5pt}
    \resizebox{\linewidth}{!}{%
    \begin{tabular}{@{}lrrrrrr@{}}
        \toprule
        \multirow{2}{*}{\centering Workload} 
        & Ref. E. & S1 E. & Flex. E. & Share & Peak flex. & Active \\
        & (MWh) & (MWh) & (MWh) & (\%) & (MW) & intervals \\
        \midrule
        Training (TR1) & 7200.00 & 7023.17 & 176.83 & 84.1 & 65.24 & 28 \\
        Inference (INF5+INF6) & 843.75 & 810.52 & 33.34 & 15.9 & 5.59 & 29 \\
        \bottomrule
    \end{tabular}%
    }
\end{table}

\vspace{-2ex}
\subsection{Checkpoint-aware Scheduling \& Monte Carlo Simulation}
With the AIDC power trajectories and checkpoint information obtained from Phase II, the grid operator can subsequently solve the checkpoint-aware two-stage robust DCOPF in Phase III. To demonstrate the advantages and necessity of the proposed coordination framework and the uncertainty modeling of checkpoint candidates, 100 scenarios were generated by Monte Carlo sampling renewable generation outputs and AIDC checkpoint events from checkpoint candidate windows. The actual operational performance of four different scheduling strategies was then evaluated and compared under these scenarios.
% Four scheduling strategies are retained in the case study:
\begin{itemize}
    \item S1: the proposed coordinated checkpoint-aware robust model;
    \item S2: the coordinated deterministic scheduling with nominal checkpoint timing (do not consider the uncertainty of checkpoint);
    \item S3: the coordinated deterministic scheduling that ignores sudden load drops induced by checkpoint;
    \item S4: a non-coordinated benchmark, where the AIDC power trajectories and checkpoint information are unavailable to the grid operator, and the AIDC operation is scheduled independently without considering grid security.
\end{itemize}

% The resulting schedules are further tested using 100 checkpoint-only Monte Carlo scenarios. Each scenario samples one timing realization from the candidate window of every checkpoint event, while renewable availability remains nominal. The first-stage schedule of each strategy is fixed and a full-day DCOPF redispatch is solved for each scenario.
The simulation results are shown in Table~\ref{tab:step3-schedule-validation}.
The proposed strategy S1 achieves zero load shedding across all tested scenarios, demonstrating the capability to maintain system security under renewable uncertainty and checkpoint uncertainty. In contrast, the non-coordinated benchmark (S4) experiences load shedding in nearly all scenarios. This is primarily because a large Training AIDC is connected at Bus 9, significantly increasing the local demand. Under periods of low renewable generation, the existing transmission capacity is insufficient to accommodate the power transfers if the AIDC operates independently of grid conditions, leading to load shedding.

\begin{table}[t]
    %\caption{Monte Carlo Simulation Comparison Result of Four Strategies}
    \caption{Performance Comparison of Four Scheduling Strategies Under Monte Carlo Simulations}
    \label{tab:step3-schedule-validation}
    \centering
    \footnotesize
    \setlength{\tabcolsep}{2.2pt}
    \resizebox{\linewidth}{!}{%
    \begin{tabular}{@{}crrrrrrr@{}}
        \toprule
        \multirow{2}{*}{Strategy}
        & Sched. & Zero
        & Avg. shed
        & Max shed
        & Avg. curt.
        & Max curt.
        & Avg. \\
        & obj. & shed & (MWh) & (MWh) & (MWh) & (MWh) & obj. \\
        \midrule
        S1 & 419,871.18 & 100/100 & 0.00 & 0.00 & 981.708 & 1023.962 & 414,849.5 \\
        S2 & 516,685.62 & 11/100 & 8.82 & 32.51 & 1047.681 & 1118.753 & 670,166.5 \\
        S3 & 511,588.55 & 92/100 & 0.36 & 6.10 & 1028.887 & 1070.700 & 579,233.3 \\
        S4 & 2,154,401.58 & 0/100 & 145.31 & 156.97 & 1026.515 & 1067.644 & 2,029,862.0 \\
        \bottomrule
    \end{tabular}
    }
\vspace{-5ex}
\end{table}

A comparison between S2 and S3 reveals an interesting observation. When checkpoint-induced power drops and rebounds are ignored (S3), the load shedding result is better than the checkpoint-aware strategy (S2). 
This is because S2 assumes that checkpoints occur at fixed time periods, which may differ from the practice. An inaccurate estimation of checkpoint timing can lead to inefficient allocation of system flexibility and reserve resources, thereby aggravating load shedding when the actual checkpoint events occur at different times.
%This is because neglecting checkpoint uncertainty effectively assumes a more favorable operating condition and therefore avoids reserving additional flexibility to accommodate potential checkpoint events. However, such schedules may become vulnerable when checkpoint realizations occur in practice. 
This result highlights the importance of explicitly modeling checkpoint uncertainty and justifies the use of the proposed robust optimization framework to ensure reliable system operation under uncertain checkpoint realizations.
% jointly reports the scheduled objective and the scenario-validation results, including load shedding and renewable curtailment. 
% The scheduled objective is the value obtained in the Step3 scheduling model, whereas the average objective is evaluated over the 100 realized checkpoint scenarios. S1 converges in five C\&CG iterations with a final upper bound of 419,871.18 and a relative gap of 0.0373\%. 

%It remains hard-feasible and zero-shed in all scenarios, showing that the checkpoint-robust schedule protects the coordinated high-utilization AIDC trajectory against timing deviations. S2 and S3 are also hard-feasible but occasionally require load-shedding recourse because their first-stage schedules do not optimize against the full checkpoint timing set. S4 sheds load in every scenario, demonstrating the operational cost of treating the Training AIDC as a static rated load without coordination or checkpoint reporting.
Although a noticeable amount of renewable curtailment is observed, the proposed strategy still achieves the lowest curtailment among all compared strategies. The curtailment is primarily caused by transmission congestion in the modified IEEE 14-bus system rather than checkpoint-induced power fluctuations, as the original transmission limits are retained, which substantially restricts the integration of large-scale renewable generation and AIDCs. In addition, renewable curtailment is only penalized in the objective function and is not explicitly forbidden in our proposed model.

\vspace{-2ex}
\subsection{NYISO Test Case}

To further validate the effectiveness and scalability of the proposed coordination framework, additional case studies are conducted on a modified NYISO test system. The same four scheduling strategies introduced above are adopted for comparison. 
The simulation results are shown in Table~\ref{tab:nyiso-validation}, including renewable curtailment and realized total operating cost over the validation scenarios.

% Table~\ref{tab:nyiso-validation} reports the scheduled objective and the resulting validation performance. 
It is observed that all four strategies remains zero load shedding in all scenarios. 
It is attributed to the abundant generation resources and transmission capacity in the original NYISO system. Specifically, the total available generation capacity is approximately 42 GW, significantly exceeding the system peak demand of 32 GW. As a result, system reliability is not the primary concern, and the key operational objective becomes the reduction of renewable energy curtailment.

It is observed that S1 and S2 achieve substantially lower renewable curtailment than S3 and S4. Unlike the modified IEEE 14-bus case, where transmission congestion and limited generation resources dominate system operation, the NYISO system possesses sufficient generation and transmission capacity. Explicitly accounting for checkpoint events (S2) yields noticeably better performance than ignoring power drops and rebounds induced by checkpoint (S3).

Comparing S1 and S2, the deterministic strategy S2 achieves a slightly lower average renewable curtailment across the 100 scenarios. It is because the proposed method is formulated as a two-stage robust optimization problem, which optimizes system performance against the worst-case realizations rather than the random cases. As a result, the robust strategy may sacrifice some average performance to enhance system resilience. Under the worst S1 C\&CG scenario, S1 results in lower renewable curtailment than S2, demonstrating the effectiveness of the proposed framework in limiting worst-case system impacts and mitigating renewable curtailment.

% Future work may explore risk-aware formulations, such as CVaR-based or distributionally robust optimization approaches, to achieve a better balance between worst-case robustness and average renewable-curtailment performance.

\begin{table}[t]
    % \caption{NYISO Checkpoint-Only Validation with C\&CG Stress Scenarios}
    \caption{Performance Comparison of Four Strategies for NYISO Testcase}
    \label{tab:nyiso-validation}
    \centering
    \footnotesize
    \setlength{\tabcolsep}{2.2pt}
    \begin{tabular}{@{}crrrrr@{}}
        \toprule
        \multirow{2}{*}{Strategy}
        & \multirow{2}{*}{Zero shed}
        & Avg. curt.
        & Avg. 
        & Max curt.
        & Max\\
        & & (MWh) & obj. & (MWh) & obj. \\
        \midrule
        S1 & 100/100 & 7.73 & 22.0468M & 40.89 & 22.0552M\\
        S2 & 100/100 & 3.79 & 21.5996M & 76.09 & 22.0777M \\
        S3 & 100/100 & 500.59 & 22.1243M & 1148.27 & 22.2538M \\
        S4 & 100/100 & 431.49 & 22.1543M & 1087.72 & 22.2854M \\
        \bottomrule
    \end{tabular}
 \vspace{-4ex}
\end{table}

\vspace{-2ex}
\section{Conclusion}
\label{sec:conclusion}

This paper presents a three-phase coordinated operation scheme between the power grid and AIDCs. 
%The first phase projects detailed grid constraints onto the AIDC power-trajectory space and constructs a V-representation inner approximation. 
%The second phase performs AI workload allocation over this certified region by jointly considering training DVFS, LT-inference DVFS, and cross-site inference workload routing. 
%The third phase validates the resulting AIDC power plan through a two-stage robust OPF with renewable and checkpoint timing uncertainties.
The proposed framework is designed to preserve both grid security and AIDC operational privacy. By exchanging only certified security-region vertices, optimized AIDC power plans, and checkpoint candidate windows, the method avoids a monolithic centralized model and reduces the need for frequent iterative communication. 
% Future work will extend the analysis to larger systems and investigate alternative robust coordination strategies under multi-operator settings and faster AIDC power fluctuations.
% The two case studies further demonstrate the necessity of grid--AIDC coordination and highlight the importance of explicitly considering checkpoint uncertainty in grid scheduling. 
The proposed framework enables AIDCs to support grid operation through workload allocation while preserving operational privacy. By exploiting the flexibility of AI workloads, the framework effectively alleviates transmission congestion, reduces load shedding and renewable curtailment, and improves overall system operating conditions. 
The results further indicate that training workloads provide substantially greater flexibility for grid support than inference workloads.
%The results further indicate that training workloads constitute the primary source of flexibility for grid support, whereas inference workloads provide comparatively limited flexibility under the considered operating conditions.
% The results show that neglecting checkpoint uncertainty may lead to inefficient resource allocation and reduced operational robustness.

% \section*{Acknowledgment}
% The authors would like to thank [to be added].

\bibliographystyle{IEEEtran}
\vspace{-2ex}
\bibliography{references}

@article{colangelo2026ai,
  title={AI data centres as grid-interactive assets},
  author={Colangelo, Philip and Coskun, Ayse K. and Megrue, Jack and Roberts, Ciaran and Sengupta, Shayan and Sivaram, Varun and Tiao, Ethan and Vijaykar, Aroon and Williams, Chris and Wilson, Daniel C. and others},
  journal={Nature Energy},
  volume={11},
  number={2},
  pages={254--261},
  year={2026},
  publisher={Nature Publishing Group UK London}
}

@article{force2025characteristics,
  title={Characteristics and risks of emerging large loads},
  author={Force, NERC Large Loads Task},
  journal={North American Electric Reliability Corporation (NERC), White Paper},
  year={2025}
}

@article{group2026assessment,
  title={Assessment of gaps in existing practices, requirements, and reliability standards for emerging large loads},
  author={Group, NERC Large Loads Working},
  journal={North American Electric Reliability Corporation (NERC), White Paper},
  year={2026}
}

@misc{meta2024llama3,
  title={The Llama 3 Herd of Models},
  author={{Llama Team, AI @ Meta}},
  year={2024},
  eprint={2407.21783},
  archivePrefix={arXiv},
  primaryClass={cs.AI},
  url={https://arxiv.org/abs/2407.21783}
}

@inproceedings{zhao2023sustainable,
  title={Sustainable supercomputing for AI: GPU power capping at HPC scale},
  author={Zhao, Dan and Samsi, Siddharth and McDonald, Joseph and Li, Baolin and Bestor, David and Jones, Michael and Tiwari, Devesh and Gadepally, Vijay},
  booktitle={Proceedings of the 2023 ACM Symposium on Cloud Computing},
  pages={588--596},
  year={2023}
}

@article{chen2020internet,
  title={Internet data center load modeling for demand response considering the coupling of multiple regulation methods},
  author={Chen, Min and Gao, Ciwei and Shahidehpour, Mohammad and Li, Zuyi and Chen, Songsong and Li, Dezhi},
  journal={IEEE Transactions on Smart Grid},
  volume={12},
  number={3},
  pages={2060--2076},
  year={2020},
  publisher={IEEE}
}

@article{radovanovic2022carbon,
  title={Carbon-aware computing for datacenters},
  author={Radovanovi{\'c}, Ana and Koningstein, Ross and Schneider, Ian and Chen, Bokan and Duarte, Alexandre and Roy, Binz and Xiao, Diyue and Haridasan, Maya and Hung, Patrick and Care, Nick and others},
  journal={IEEE Transactions on Power Systems},
  volume={38},
  number={2},
  pages={1270--1280},
  year={2022},
  publisher={IEEE}
}

@article{hall2025carbon,
  title={Carbon-aware computing for data centers with probabilistic performance guarantees},
  author={Hall, Sophie and Micheli, Francesco and Belgioioso, Giuseppe and Radovanovi{\'c}, Ana and D{\"o}rfler, Florian},
  journal={IEEE Transactions on Power Systems},
  year={2025},
  publisher={IEEE}
}

@article{wu2025game,
  title={Game-Based Optimization Method for Geo-Distributed Data Centers Under Customer Directrix Load Demand Response Mechanism},
  author={Wu, Jia-Kai and Liu, Zhi-Wei and Zhao, Yong and Li, Wenqu and Li, Yuanzheng},
  journal={IEEE Transactions on Smart Grid},
  year={2025},
  publisher={IEEE}
}

@article{dvorkin2024agent,
  title={Agent coordination via contextual regression (agentconcur) for data center flexibility},
  author={Dvorkin, Vladimir},
  journal={IEEE Transactions on Power Systems},
  volume={40},
  number={2},
  pages={1832--1842},
  year={2024},
  publisher={IEEE}
}

@article{chen2025spatial,
  title={Spatial flexibility provision from geographically dispersed data centers enabling coordinated operation of multi-local flexibility markets},
  author={Chen, Boyu and Che, Yanbo and Takci, Mehmet and Qadrdan, Meysam and Zhou, Yue},
  journal={IEEE Transactions on Smart Grid},
  year={2025},
  publisher={IEEE}
}

@article{wang2025multi,
  title={Multi-Objective Low-Carbon Scheduling Method for Data Centers Based on Ensemble Reinforcement Learning},
  author={Wang, Yifan and Sun, Wei and Ren, Pengyu and Harrison, Gareth},
  journal={IEEE Transactions on Smart Grid},
  year={2025},
  publisher={IEEE}
}

@article{al2021potential,
  title={Potential of data centers for fast frequency response services in synchronously isolated power systems},
  author={Al Kez, Dlzar and Foley, Aoife M and Ahmed, Faraedoon W and O'Malley, Mark and Muyeen, SM},
  journal={Renewable and Sustainable Energy Reviews},
  volume={151},
  pages={111547},
  year={2021},
  publisher={Elsevier}
}

@article{zou2025coordinating,
  title={Coordinating multiple geo-distributed data centers for enhanced participation in frequency regulation services under uncertainty},
  author={Zou, Bin and Chen, Ge and Zhang, Hongcai and Song, Yonghua},
  journal={Journal of Modern Power Systems and Clean Energy},
  year={2025},
  publisher={SGEPRI}
}

@article{ren2026grid,
  title={Grid frequency stability support potential of data center: A quantitative assessment of flexibility},
  author={Ren, Pengyu and Sun, Wei and Wang, Yifan and Harrison, Gareth},
  journal={IEEE Transactions on Industry Applications},
  year={2026},
  publisher={IEEE}
}

@article{chen2025electricity,
  title={Electricity demand and grid impacts of AI data centers: Challenges and prospects},
  author={Chen, Xin and Wang, Xiaoyang and Colacelli, Ana and Lee, Matt and Xie, Le},
  journal={arXiv preprint arXiv:2509.07218},
  year={2025}
}

@article{kakolyris2024slo,
  title={Slo-aware gpu dvfs for energy-efficient llm inference serving},
  author={Kakolyris, Andreas Kosmas and Masouros, Dimosthenis and Xydis, Sotirios and Soudris, Dimitrios},
  journal={IEEE Computer Architecture Letters},
  volume={23},
  number={2},
  pages={150--153},
  year={2024},
  publisher={IEEE}
}

@inproceedings{kakolyris2025throttll,
  title={throttll’em: Predictive gpu throttling for energy efficient llm inference serving},
  author={Kakolyris, Andreas Kosmas and Masouros, Dimosthenis and Vavaroutsos, Petros and Xydis, Sotirios and Soudris, Dimitrios},
  booktitle={2025 IEEE International Symposium on High Performance Computer Architecture (HPCA)},
  pages={1363--1378},
  year={2025},
  organization={IEEE}
}

@article{he2025freesh,
  title={FREESH: Fair, Resource-and Energy-Efficient Scheduling for LLM Serving on Heterogeneous GPUs},
  author={He, Xuan and Fang, Zequan and Lian, Jinzhao and Tsang, Danny HK and Zhang, Baosen and Chen, Yize},
  journal={arXiv preprint arXiv:2511.00807},
  year={2025}
}

@article{lohne2016equivalence,
  title={Equivalence between polyhedral projection, multiple objective linear programming and vector linear programming},
  author={L{\"o}hne, Andreas and Wei{\ss}ing, Benjamin},
  journal={Mathematical Methods of Operations Research},
  volume={84},
  number={2},
  pages={411--426},
  year={2016},
  publisher={Springer}
}

@article{fischetti2009minimal,
  title={Minimal infeasible subsystems and Benders cuts},
  author={Fischetti, Matteo and Salvagnin, Domenico and Zanette, Arrigo and others},
  journal={Mathematical Programming to appear},
  year={2009}
}

@misc{liu_parameters_ieee14_nyiso,
  author       = {Ziang Liu},
  title        = {parameters\_for\_ieee14\_and\_NYISO},
  year         = {2026},
  howpublished = {\url{https://github.com/TsMikeLiu/parameters_for_ieee14_and_NYISO}},
  note         = {GitHub repository, accessed 2026-06-01}
}

@article{liu2025synergising,
  title={Synergising Hierarchical Data Centers and Power Networks: A Privacy-Preserving Approach},
  author={Liu, Junhong and Teng, Fei and Hou, Francis Yunhe},
  journal={IEEE Transactions on Smart Grid},
  volume={16},
  number={6},
  pages={5083--5098},
  year={2025},
  doi={10.1109/TSG.2025.3603107}
}

\end{document}